\documentclass[a4paper,11pt]{article}
\pdfoutput=1 
\usepackage{jcappub} 
\usepackage[T1]{fontenc}
\usepackage[british]{babel}
\usepackage{xcolor}
\usepackage{booktabs}
\usepackage{latexsym}
\usepackage{amssymb}
\usepackage{amsmath}
\usepackage{graphicx}
\usepackage{tabularx}
\usepackage{cancel}
\usepackage{hyperref}
\usepackage{ulem}
\usepackage[]{cleveref}
\crefname{plural}{Eqs.}{Eqs.}
\Crefname{plural}{Eqs.}{Eqs.}

\crefformat{plural}{#2eqs.~(#1)#3}
\Crefformat{plural}{#2Eqs.~(#1)#3}

\newcommand{\dd}{\text{d}}
\newcommand{\aap}{Astron. \& Astrop.}

\title{\boldmath Cosmological coupling of local gravitational systems}
\author[a,b]{Mariano Cadoni,}
\author[a,b]{ Mirko Pitzalis,}
\author[c]{Davi C. Rodrigues,}
\author[a,b]{ Andrea P. Sanna}

\affiliation[a]{Dipartimento di Fisica, Universit\`a di Cagliari, Cittadella Universitaria, 09042 Monserrato, Italy}
\affiliation[b]{INFN, Sezione di Cagliari, Cittadella Universitaria, 09042 Monserrato, Italy}
\affiliation[c]{Departamento de F\'isica \& N\'ucleo Cosmo-Ufes, Universidade Federal do Esp\'irito Santo, 29075-910 Vit\'oria, ES, Brazil}

\emailAdd{mariano.cadoni@ca.infn.it}
\emailAdd{mirko.pitzalis@ca.infn.it}
\emailAdd{davi.rodrigues@ufes.br}
\emailAdd{asanna@dsf.unica.it}

\abstract{
We investigate the cosmological coupling of  spherical, local  astrophysical systems.
We derive a general formula quantifying the cosmological coupling of the Misner-Sharp mass of these objects. We show that, in the weak-field limit, the cosmological coupling is only allowed if there are pressure anisotropies. We also apply our results to galaxies, modelling them with the Navarro-Frenk-White and Einasto profiles. We show that the galactic mass can be coupled to the cosmological dynamics and examine its dependence on the scale factor of the universe.
}

\begin{document}
\maketitle
\flushbottom
\section{Introduction}

The possibility of a dynamical coupling between local astrophysical objects like, e.g., black holes or stars, and the large-scale cosmological dynamics in a General Relativity (GR) framework has a quite long and partially contradictory history (see, e.g., Refs.~\cite{Gao:2023keg,McVittie:1933zz,Einstein:1945id,Einstein:1946zz,Pachner:1963zz,dicke1964evolution,Vaidya:1968zza,DEath:1975jps,Gautreau:1984pny,Cooperstock:1998ny,Nayak:2000mr,Baker:2000yh,Bolen:2000dz,Dominguez:2001it,Ellis:2001cq,Gao:2004cr,Sheehan:2004wa,Nesseris:2004uj,Sultana:2005tp,Li:2006zh,Adkins:2006kw,McClure:2006kg,Sereno:2007tt,Faraoni:2007es,Balaguera-Antolinez:2007csw,Mashhoon:2007qm,Carrera:2008pi, Gao:2011tq,Faraoni:2014nba,Kopeikin:2014qna,Faraoni:2015saa,Mello:2016irl,Faraoni:2018xwo,Guariento:2019ock,Spengler:2021vxy,Agatsuma:2022ewd}). 
In the last few years, there have been rather interesting developments on the issue, both from the theoretical and observational point of view. 

From the theoretical side, two different approaches have emerged. 
One employs perturbations and an averaging procedure \cite{Croker:2019mup,Croker:2020,Croker:2020plg} finding that the coupling can be traced back to pressures within local compact astrophysical objects. This effect manifests itself as a power-law relation between the local object masses and the scale factor $a$, i.e., $M(a) \propto a^k$~\cite{Croker:2019mup}, with $-3\le k\le 3$.
These results and the consequent claim that black holes might be the source of dark energy \cite{Farrah:2023opk} have been criticised from different points of view. First, the framework adopted could be flawed from the beginning~\cite{Mistele:2023fds}. Second, the huge separation of scales between local objects and the cosmological dynamics should make such coupling physically implausible~\cite{Wang:2023aqe,Gaur:2023hmk}. Third, the equation of state of matter inside a black hole could hardly mimic dark energy~\cite{Parnovsky:2023wkc,Avelino:2023rac,Dahal:2023hzo}. 

The other approach tackles the cosmological coupling problem using a solid GR framework \cite{Cadoni:2023lum, Cadoni:2023lqe}, along the lines of the first work on the subject by McVittie~\cite{McVittie:1933zz}. It is based on a general cosmological embedding of local objects and gives a general description of the coupling between local inhomogeneities and the cosmological background. The origin of the cosmological coupling is here traced back to a curvature term in Einstein's equations, giving the precise prediction $M(a) \propto a^k$ with $k=1$. Moreover, it was shown that the cosmological coupling is not generic for compact objects, but limited to specific ones characterised by a mismatch between their ADM and Misner-Sharp (MS) masses \cite{Cadoni:2023lqe}. 
A striking consequence of this result is that, while usual singular black holes do not couple to the cosmological background, non-singular ones do. Thus, an observational evidence of a cosmological coupling of astrophysical black holes with coupling exponent $k=1$ would be the smoking gun of their non-singular nature \cite{Cadoni:2023lqe}.

From the observational side, the situation is quite intricate. The first test of the cosmological mass shift formula was performed in Ref.~\cite{Farrah:2023opk} using a sample of supermassive black holes at the centre of elliptical galaxies at different redshifts. This set of data showed a preference for $k \sim 3$~\cite{Farrah:2023opk,Cadoni:2023lum}, which eventually motivated the speculative connection between dark energy and black holes. 
Other observational constraints on the scaling parameter $k$ are rather controversial, as they seem to strongly depend on the astrophysical sample employed in the analysis~\cite{Rodriguez:2023gaa,Andrae:2023wge,Lei:2023mke,Amendola:2023ays,Lacy:2023kbb}.

Leaving aside these issues on the observational side, the main goal of this paper is to generalize the results of Refs.~\cite{Cadoni:2023lum, Cadoni:2023lqe} to generic, local, static and spherically-symmetric, astrophysical objects, in particular those in the weak-field regime. 
Indeed, a limitation of the results of Refs.~\cite{Cadoni:2023lum, Cadoni:2023lqe} is that the cosmological mass shift formula derived there applies only to strongly-coupled, compact objects, characterized by a linear relation between their mass $M$ and radius $L$. Therefore, it is of interest to generalize these results to generic local objects, without any assumption on their compactness.
Another issue is related to understanding how specific is the embedding of astrophysical bodies in a Friedmann-Lema\^{i}tre-Robertson-Walker (FLRW) cosmological background. Specifically, one would like to understand whether generic local gravitational systems, especially Newtonian ones, allow for cosmological embedding. Answering this question is crucial to shed light on the possible cosmological coupling of astrophysical systems of interest, such as galaxies, galaxy clusters or binary systems, which are known to be in the weak-field regime. Addressing the above issues is further complicated by the fact that the existence of cosmologically-embedded solutions typically requires an anisotropic fluid as local source. Thus, the weak-field limit of gravitational systems characterized by this kind of sources needs to be kept under control.

In this paper, we address these issues building on the results of Refs.~\cite{Cadoni:2023lum, Cadoni:2023lqe}. We first generalize the mass formula of Ref.~\cite{Cadoni:2023lum} to generic spherically-symmetric, local 
astrophysical objects (\cref{sect:cc}). Then, we delve into the weak-field limit of these solutions sourced by an anisotropic fluid and discuss their cosmological embedding in a FLRW background (\cref{sect:ce}). Consequently, we apply our results to the cosmological coupling of galaxies (\cref{sect:galaxies}). Lastly, we draw our conclusions in \cref{sec:Conclusions}.

\section{Cosmological coupling of generic local astrophysical objects}
\label{sect:cc}

In Refs.~\cite{Cadoni:2023lum,Cadoni:2023lqe}, a mass formula describing the cosmological coupling of strongly-coupled, ultra-compact astrophysical objects was derived. A crucial assumption in the derivation was the linear scaling of the  mass of the spherically-symmetric object with its radius 
\cite{Cadoni:2023lqe}. This is surely the case for black holes or other compact objects, but not for other astrophysical bodies, such as stars, galaxies, or galaxy clusters.
From general arguments, we expect the mass of any local, spherically-symmetric solution of GR, allowing for a cosmological embedding, to depend on the redshift.

In this section, we will generalize the mass formula of Refs.~\cite{Cadoni:2023lum,Cadoni:2023lqe} to generic, spherically-symmetric objects that allow for a cosmological embedding. Throughout this paper, we will consider the cosmological background as fixed, i.e., fully determined by the usual FLRW cosmology. This is because we are only interested in the cosmological embedding of local structures in a regime sufficiently far away from the transition scale to the large-scale homogeneity and isotropy.
Therefore, throughout the paper, we will assume that such embedding is permitted. This assumption is rather strong, as it is not allowed by every static, spherically-symmetric solution of GR \cite{Cadoni:2020jxe}. Additionally, we assume that the object is locally static on the time-scale of the cosmological evolution. In other words, the only time-dependence comes from its cosmological coupling. This implies the exclusion of any other local dynamical process, such as astrophysical accretion mechanisms. 

Following Refs.~\cite{Cadoni:2023lum,Cadoni:2023lqe}, the spacetime is described by the metric
\begin{equation}
\dd s^2 = a^2(\eta)\left[-e^{\alpha(\eta, r)} \dd \eta^2 + e^{\beta(\eta, r)} \dd r^2 + r^2 \dd\Omega^2\right]\, ,
\label{generalmetric}
\end{equation}
where $\eta$ is the conformal time.

We consider the general situation in which the local object is sourced by an anisotropic fluid, characterized by radial and tangential pressures $p_{\parallel}$, $p_\perp$, respectively. The usual isotropic fluid is recovered as a particular case by setting $p_{\parallel}=p_\perp\equiv p$.

The resulting Einstein's equations allow for two branches of solutions (for details, see Ref.~\cite{Cadoni:2023lum}). For our purposes, the interesting solutions are those characterized by $\dot \alpha=0$. In this case, the field equations and stress-energy tensor conservation take the form \cite{Cadoni:2023lum,Cadoni:2023lqe} 
\begin{subequations}
\begin{align}
& e^{-\beta(r, \eta)} = e^{-\beta_0(r)} \left(\frac{a(\eta)}{a(\eta_0)}\right)^{k(r)}\, ;
 \label{betafunction}\\
&\frac{\dot{a}^2}{a^2}\left(3-r\alpha' \right)e^{-\alpha} + \frac{1-e^{-\beta}+r\beta'e^{-\beta}}{r^2} = 8\pi Ga^2 \rho\, ; \label{alpha00}\\
&\frac{e^{-\beta}+re^{-\beta} 	\alpha'-1}{r^2}+e^{-\alpha}\left(-2\frac{\ddot{a}}{a}+\frac{\dot{a}^2}{a^2} \right)=8\pi G a^2 p_{\parallel}\, ; \label{alpharr}\\
&p'_{\parallel} +\frac{\alpha'}{2}\left(\rho+p_{\parallel}\right)+ \frac{2}{r}\left(p_{\parallel}-p_\perp\right)=0 \label{pconserv}\, ,
\end{align}
\label{systemalphadotzero}
\end{subequations}
where 
\begin{equation}\label{kr}
k(r) \equiv r \alpha' (r)\, ,
\end{equation}
$\eta_0$ is a reference time, whereas $\beta_0(r)$ is the metric function evaluated at the latter. $\beta_0(r)$ represents more than a simple initial condition. We are assuming that local profiles of the objects remain the same throughout the cosmological evolution. Physically, this is consistent with our assumption of neglecting any other local, time-dependent dynamics of the  object. The latter is then assumed to be static. The only permitted time-dependence comes from the cosmological expansion, represented by the factorization in \cref{betafunction}. 
Thus, $\beta_0(r)$ gives the static profile of the local solution at every value of the cosmological time. Notice that the factorized form \eqref{betafunction} is compatible with \cref{alpha00,alpharr,pconserv} since the system is not closed. The latter equations have to be considered as the definitions of the $\eta$- and $r$-dependent density and pressures. We still have the freedom to choose the reference time $\eta_0$ and the normalization of the scale factor at that time. Throughout this paper, we will choose this reference time as the present one and adopt the standard scale factor normalization $a(\eta_{0}) = a_0 = 1$.

Before solving the relevant time-dependent equations, we consider a simpler problem. At $\eta_0$, assuming that $\dot a$ and $\ddot a$ are sufficiently small, \cref{alpha00,alpharr,pconserv} give
\begin{subequations}
\label{SystemEqFirstRegime}
\begin{align}
&\frac{1-e^{-\beta_0}+r\beta_0'e^{-\beta_0}}{r^2} = 8\pi G \, \rho(r)\, ; \label{alpha00static}\\
&\frac{e^{-\beta_0}+re^{-\beta_0} 	\alpha'-1}{r^2}=8\pi G  \, p_{\parallel}(r)\, ; \label{alpharrstatic} \\
&p'_{\parallel} +\frac{\alpha'}{2}\left(\rho+p_{\parallel}\right)+ \frac{2}{r}\left(p_{\parallel}-p_\perp\right)=0 \label{press1}\, .
\end{align}
\end{subequations}
In the above, $\rho(r)$, $p_{\parallel}(r)$ are the time-independent density and pressure profiles of the local object at the present time. 

One can easily integrate \cref{alpha00static} to obtain 
\begin{equation}\label{beta0solutiongeneral}
e^{-\beta_0(r)} = 1-\frac{2Gm(r)}{r}\, , \qquad m(r) = 4\pi  \, \int^r \dd \tilde r \, \tilde r^2 \rho(\tilde r)\, .
\end{equation}
Then one can solve \cref{alpharrstatic,press1} for $\alpha(r)$.

Once $\beta_0$ is determined, the full cosmologically-embedded solution is obtained by using \cref{betafunction} to get $\beta(\eta,r)$. The time-dependent pressures and density $p_{\parallel}(\eta,r)$, $\rho(\eta,r)$, sourcing the cosmologically coupled solution, are then given by \cref{alpha00,alpharr}.

As a matter of fact, the existence of the time-dependent solution is generally only guaranteed if the tangential pressure $p_\perp$ is not fixed, i.e., given by a free function. Conversely, if $p_{\perp}(r)$ is fixed, it may happen that the conservation equation \eqref{pconserv} gives an independent equation, the full system becomes over-constrained and the cosmologically-embedded solution does not exist. We will see in the next section that this is crucial for the existence of cosmologically coupled solutions describing local gravitational objects in the weak-field regime.  

The MS mass of the cosmologically-coupled solution can be inferred from the integration of \cref{alpha00} over the proper volume (see \cite{Cadoni:2023lum,Cadoni:2023lqe}). Neglecting the cosmological density term, which is assumed to be negligible with respect to the density of the local object, we get the mass shift between some arbitrary initial $\eta_i$ and final $\eta_f$ cosmological times 

\begin{equation}\label{massshift}
M(\eta_f,L)= M(\eta_i,L)\frac{a_f}{a_i} \frac{ 1-e^{-\beta_0(L)}a_f^{k(L)}}{ 1-e^{-\beta_0(L)}a_i^{k(L)}}\,, 
\end{equation}
where $L$ is the radius (or size) of the local object. For black holes, this is identified as the event horizon radius. For horizonless objects, $L$ is defined as the radius of the sphere that contains $99\%$ of the object's mass \cite{Cadoni:2023lum}.

To ease the notation, in the following, we use $k(L) \equiv k_L$ and $\beta_0(L)\equiv \beta_0$. \Cref{massshift} represents the most general expression describing the cosmological redshifting of the mass of generic, comologically-coupled local objects. 

The mass formula obtained in  \cite{Cadoni:2023lum,Cadoni:2023lqe} can be recovered applying \cref{massshift} to very compact objects, for which the initial mass satisfies $M\propto a_i L$. If the object is endowed with an event horizon, we get a pure linear scaling of $M(a)$
\begin{equation}\label{lscal}
M(\eta_f,L)= M(\eta_i,L)\frac{a_f}{a_i}\, ,
\end{equation}
which is also a particular case of Croker's scaling law \cite{Croker:2019mup,Croker:2020,Croker:2021duf,Croker:2020plg,Croker:2019kje}. Indeed, let  $r_{h}$ be the horizon radius, thus in the limit $L \to r_{h}$, one finds $e^{- \beta_0(L)} a^{k_L} \to 0$. The same behavior of \cref{lscal} is also recovered from \cref{massshift} whenever $k_L = 0$ and regardless of the presence of horizons.

For generic objects allowing for cosmological embedding, instead, \cref{massshift} holds in its generality, giving also a sublinear behavior. In this case, the linear behavior \eqref{lscal} holds only approximately near $a_f\sim a_i$. Indeed, by expanding \cref{massshift} near this point, we get \cref{lscal} as leading term.

\subsection{Quasi-asymptotic flatness and behavior of the mass formula}
\label{subsec:behaviormassformula}

At fixed time and at distances much greater than the object radius $L$, but much smaller than the Hubble radius, the static gravitational field of any local, isolate, gravitational system must decay to zero similarly to the Newtonian behavior. This simple physical feature constrains the behavior of the mass formula \eqref{massshift}, resulting in $\beta_0 > 0$. We write the mass $M$ as a function of $a$, for $ 0\le a\le 1$,
\begin{equation}\label{GeneralMassFormula}
M(a)= \frac{L a}{2G} \left( 1-e^{-\beta_0} a^{k_L}\right)\, . 
\end{equation}
$M(a)$ can behave in two possible ways:
\begin{enumerate}
    \item If $k_L$ is close to zero, it increases monotonically from $0$ to $M(1)$;
    \item If $k_L$ is large, it starts increasing  linearly near $a=0$, thus reaching a maximum at $a_M$, and then it goes down to $M(1)$. 
\end{enumerate}

The presence of the maximum can be detected by looking for the zeroes of the first derivative of $M(a)$. One finds
\begin{equation}\label{max}
k_L\ln a_M= -  \ln \left[e^{-\beta_0}(k_L+1)\right]\, .
\end{equation}
For $a < 1$, this equation admits a solution only for 
\begin{equation}\label{ineq}
e^{-\beta_0}\ge \frac{1}{k_L+1} \,.
\end{equation}
Whenever this inequality is satisfied, case $2$ is realized. 

The above analysis shows the non-trivial evolution of the quasi-local mass and its dependence on the cosmological background. Considering the physical origin of the cosmological coupling of local structures, this behavior can be traced back to the local curvature and these results might, therefore, be interpreted as follows. On the one hand, if \cref{ineq} is satisfied, the local structure becomes weakly coupled at $r=L$, leading to a decrease in mass when $a$ increases. On the other hand, if \cref{ineq} is violated, the physical system is still in a strong-coupling regime, in a similar manner to the black-hole case, implying the increase of $M$ with $a$.

\section{Cosmological embedding in the weak-field limit}
\label{sect:ce}

In the previous section, we derived the MS mass for a generic local, spherically-symmetric object, assuming that a GR solution describing its cosmological embedding exists. The existence of the latter is a quite involved issue, as there are no general theorems ensuring that a static, spherically-symmetric solution of GR can be promoted to the form \eqref{generalmetric}.
Nevertheless, cosmological embedded solutions are known in a number of cases, including the Schwarzschild black hole \cite{McVittie:1933zz} and regular black holes \cite{Cadoni:2023lum, Cadoni:2022chn}.

Most of the objects considered so far are gravitationally strongly-coupled, i.e., they cannot be described using the weak-field approximation at scales of the order of their radius $L$.
In this section, we will discuss the cosmological embedding of local objects in the weak-field regime. Among others we have galaxies and galaxy clusters.

The investigation of the cosmological embedding in the weak-field limit is also important from an alternative standpoint. At great distances $r\gg L$, even strongly-coupled objects allow for a weak-field description in GR. Therefore, the cosmological embedding in the weak-field approximation is expected to provide insights about known results on the cosmological embedding of black holes \cite{Cadoni:2023lqe}.

We start our investigation by first considering the weak-field limit of the field equations at constant time \eqref{SystemEqFirstRegime}. 

The weak-field limit alone is not sufficient for recovering the Newtonian limit within GR. In particular, the latter requires the pressures (or any anisotropic stress) to be sufficiently small, so as not to influence the metric solution directly. If the pressures are not sufficiently small to be compatible with the Newtonian limit, the potential associated to $g_{00}$ does not need to satisfy a Poisson equation. Therefore, we deal with three cases of weak-field limit:
\begin{enumerate}
    \item {\sl{Newtonian}}: the weak-field gravitational potential $\phi$ satisfies the standard Poisson equation and the pressures are negligible (Newtonian fluid limit), i.e., $p_{\parallel}(r)=p_\perp(r)=0$;
    \item 
    {\sl{Non-Newtonian}}, which includes two subcases: \\
    $2.a\,$ {\sl{Poissonian}}: $\phi$ still satisfies the Poisson equation, but some components of the pressure are not zero; \\
    $2.b\,$ {\sl{Non-Poissonian}}: $\phi$ does not satisfy the Poisson equation in its standard form.
\end{enumerate}
We will show below that any isotropic Poissonian weak-field limit is necessarily Newtonian. Notice also that case $2.a$ implies that the far-field generated by a gravitational system decays as $1/r$ for $r\to\infty$, but anisotropies in the pressures are necessary. This is, for instance, the case of regular black holes with de Sitter core discussed in \cite{Cadoni:2023lum, Cadoni:2022chn}. Schematically, we will show that 
  \begin{subequations}
  \begin{align}
  \mbox{Weak field limit } \xrightarrow{p_\parallel \approx-2p_\perp } \mbox{ Poissonian } \xrightarrow{p_\parallel \approx p_\perp \approx 0} \mbox{ Newtonian }; \, \\
  \mbox{Weak field limit } \xrightarrow{p_\parallel \neq-2p_\perp  } \mbox{ non-Poissonian }.            \quad\quad \quad\quad\quad\quad\quad\,\,\,\,\, 
  \end{align}
  \end{subequations}
To investigate the weak-field limit in the static case, we perturb the metric functions around the Minkowski background
\begin{subequations}
\begin{align}
    e^{\alpha} & \simeq 1+ 2 \phi(r) + \mathcal{O}(\phi^{2})\\
    e^{-\beta_0}  & \simeq 1+ 2 \psi(r) + \mathcal{O}(\psi^{2})
\end{align}
\end{subequations}
where $\phi(r)$ and $\psi(r)$ are two potentials (the former is the Newtonian one). We expand all quantities up to the linear order $\mathcal{O}(\phi)$ and $\mathcal{O}(\psi)$. Therefore
\begin{subequations}
    \begin{align}
    \alpha & = \ln(1+2\phi) \simeq 2 \phi + \mathcal{O}(\phi^2)\, ;\\
    \beta_0 & = - \ln(1+2\psi) \simeq -2 \psi + \mathcal{O}(\psi^2)\, . 
    \end{align}
\end{subequations}
At linear order, \cref{alpha00static,alpharrstatic} become 
\begin{subequations}
\begin{align}
    4\pi G \rho  &\simeq - \frac{1}{r^{2}} \partial_{r} \left( \psi r \right)\, ; \label{rhoWFL}\\
    4\pi G p_{\parallel} &\simeq \left( \frac{\psi}{r^{2}} + \frac{\phi'}{r}\right) \label{pparWFL}\, .
\end{align}
\end{subequations}
The conservation equation \eqref{press1} finally reads
\begin{equation}\label{consWFL}
   \left(r^2 p_{\parallel}\right)'\simeq \frac{1}{ 4\pi G } \left(\psi' + r\phi''+\phi'\right) \simeq 2 rp_\perp \, ,
\end{equation}
where the higher-order terms, originating from $\alpha' (\rho + p_\parallel)$, have been neglected. Let us now discuss separately the different weak-field regimes.

\subsection{Newtonian and Poissonian weak-field limits}
The Poisson equation holds in its usual form in both the Poissonian and Newtonian weak-field limits: 
\begin{equation}
    \nabla^{2} \phi = 4 \pi G \rho\, . 
    \label{PoissonEq}
\end{equation}
As a result, we can classify the weak-field limit either as Newtonian or Poissonian depending on the pressures $p_\parallel(r)$, $p_\perp(r)$. 
Comparing \cref{rhoWFL,PoissonEq} yields
\begin{equation}
    \nabla^{2} \phi = \frac{1}{r^{2}} \partial_{r} (r^{2} \phi^{\prime}) = - \frac{1}{r^{2}} \partial_{r} \left( \psi r \right)\, ,
\end{equation}
which implies the following relation between the potentials
\begin{equation}
    \psi = \frac{c_1}{r} - r\phi^{\prime}\, ,
\label{psiphirelation}
\end{equation}
with $c_1$ an integration constant. Accordingly, the pressure $p_{\parallel}$ is
\begin{equation}{\label{pparallelNewtoniansolution}}
    p_{\parallel}= \frac{c_1}{4 \pi G r^{3}}\, .
\end{equation}
From \cref{consWFL} we finally get
\begin{equation}\label{consfinalWFL}
  p_\perp\simeq -  \frac{c_1}{8 \pi G r^{3}}= -\frac{1}{2}  p_{\parallel} \, .
\end{equation}
Depending on the value of the integration constant $c_1$, we can have two subcases.

\subsubsection{Case $c_1=0$: Newtonian weak-field limit}
From \cref{pparallelNewtoniansolution,consfinalWFL}, we see that this case corresponds to a Newtonian fluid, in which the pressures are negligible, i.e., $p_{\parallel} = p_{\perp} = 0$. 

The remaining equations are
\begin{subequations}
\begin{align}
    \psi &= - r\phi^{\prime}\, ;\label{psisolc10}\\
    \rho &= \frac{1}{4 \pi G r^{2}} \left( r^{2} \phi^{\prime} \right)^{\prime}\, .
\end{align}
\end{subequations}
To close the system, the density profile must be specified. We note that, when $\rho = 0$ (the point-like source case), $\phi = -\psi = -c_2/r$ (with $c_2$ another integration constant), which is the standard Newtonian contribution.

\subsubsection{Case $c_1 \neq 0$: Poissonian weak-field limit}

In this case, the pressures are not zero and we have 
\begin{subequations}
\begin{align}\label{pwfl}
    \psi &= \frac{c_1}{r}- r\phi^{\prime}\, ;\\
    \rho &= \frac{1}{4 \pi G r^{2}} \left( r^{2} \phi^{\prime} \right)^{\prime}\, ,
\end{align}
\end{subequations}
whereas the pressure components must satisfy
\begin{equation}\label{cons}
    p_\parallel+2p_{\perp}=0\, .
\end{equation}
Also here the system must be closed by specifying a density profile $\rho(r)$. An interesting example of a Poissonian weak-field limit, which will also be considered when we will discuss the case of galaxies, is recovered by choosing a logarithmic gravitational potential
\begin{equation}\label{phic1neq0}
 \phi = A\ln\left(\frac{r}{\delta}\right)\, ,
\end{equation}
where $A$ and $\delta$ are some constants. \Cref{pwfl} gives
\begin{equation}\label{psic1neq0}
    \psi = -\left( A- \frac{c_1}{r} \right)\, ,
\end{equation}
while, from the Poisson equation, the density profile reads as
\begin{equation}\label{gh}
    \rho = \frac{A}{4 \pi G r^{2}}\, .
\end{equation}
From \cref{consfinalWFL} the pressures are 
\begin{equation}
\label{pperpNewtoniansolution}
   p_{\perp}=-\frac{1}{2} p_{\parallel}= \frac{c_1}{4 \pi G r^{3}}\, .
\end{equation}
Summarizing, for a generic anisotropic fluid in the weak-field limit, if we impose the Poisson equation, we can have: $(1)$ the usual Newtonian limit, when $p_{\perp}=p_{\parallel}=0$; $(2)$ a non-Newtonian, but still Poissonian, weak-field limit, in which both pressures do not vanish and satisfy \cref{cons}.

It is important to stress that \cref{cons} forbids any isotropic Poissonian weak-field limit different from the Newtonian one $p_{\perp} = p_{\parallel} = 0$.  

\subsection{Non-Poissonian weak-field limit}
If we do not impose the validity of the Poisson equation in the form of \cref{PoissonEq}, the weak-field approximation gives
\begin{equation}
    \nabla^{2} \phi =  4 \pi G \left(\rho + p_{\parallel}+2p_{\perp}\right) \, ,
    \label{PoissonEqmod}
\end{equation}
i.e., the Poisson equation is sourced by the active mass and we have a generic non-Poissonian weak-field limit.
Notice that the Poissonian and Newtonian weak-field limits can be obtained as particular cases of \cref{PoissonEqmod}, by setting $p_{\parallel}+2p_{\perp}=0$ or $p_{\parallel}=p_{\perp}=0$, respectively. 

\subsection{Cosmological embedding }

Having under control the weak-field limit of static, spherically-symmetric solutions, we are ready to embed these solutions into a fixed FLRW cosmological background.   
In order to do this, we will use a perturbative expansion of the $r-$dependent part of the metric functions $\beta$ and $\alpha$ into the system \eqref{systemalphadotzero}. 

This approach relies on the possibility of separating the space- and time-dependent components of the metric functions. Actually, in the case under consideration, this is indeed possible since the metric function $\alpha$ does not depend on time, whereas the time dependence of $\beta$ occurs in the simple functional form \eqref{betafunction}.

Notice also that, in our perturbative approach, we will completely neglect the gauge fixing issues. This approach is expected to be consistent as long as we are only interested in the existence of cosmologically embedded solutions.

By virtue of \cref{betafunction}, the first order in the perturbations gives  
\begin{equation}
\begin{split}
    e^{-\beta(r, \eta)} &\simeq 1 + 2 r \phi' \, \ln a + 2\psi\,,\\
    \beta(r, \eta) &\simeq -2\psi - 2 r \phi' \, \ln a\, ,\\
    \beta' &\simeq -2\psi'-2r\phi'' \, \ln a - 2\phi' \, \ln a\,,\\
    e^{-\alpha}&  \simeq 1-2\phi\, .
    \end{split}
\label{betaexpansioncosmological}
\end{equation}
By inserting these expansions into \cref{alpha00,alpharr}, we obtain 
\begin{subequations}
\begin{align}
    &\frac{\dot a^2}{a^2} \left(3-2r\phi' - 6\phi \right) - \frac{2\psi + 4 r \phi' \ln a+ 2r \psi' + 2r^2 \phi'' \ln a}{r^2}\simeq 8\pi G a^2 \rho \, ,\label{hh1}\\
    &\frac{2r \phi' \ln a + 2\psi + 2 r \phi'}{r^2} + (1-2\phi) \left(\frac{\dot a^2}{a^2}-2\frac{\ddot a}{a}\right) \simeq 8\pi G a^2 p_\parallel\, \label{hh2}.
\end{align}
\end{subequations}
Let us first consider a static, isotropic, Newtonian source, for which $p_\parallel(r) \simeq p_\perp(r) \simeq 0$. 
The cosmological uplifting of this local solution can be performed in two different ways: $1.)$ by assuming a fully isotropic source at all scales, i.e., $p_\parallel(r, \eta) = p_\perp(r, \eta)$ everywhere; $2.) $ by allowing for $p_\parallel(r, \eta) \neq p_\perp(r, \eta)$. 
This second possibility introduces the difficulty of explaining the physical mechanism producing anisotropies from isotropic local solutions, which then average again on large cosmological scales to isotropic configurations, giving rise to the usual FLRW model. Currently, we are not interested in describing these transition phenomena. In this section, we will only consider the fully isotropic case $1.)$.  In the next section we will discuss solutions with small anisotropies in the  pressures. 

Using \cref{psisolc10}, \cref{hh1,hh2} become 
\begin{subequations}
\begin{align}
    &\frac{\dot a^2}{a^2} \left(3-2r\phi' - 6\phi \right) + \frac{2\phi'}{r}-\frac{4\phi'}{r}\ln a+\frac{2}{r}(\phi'+r\phi'')-2\phi''\ln a \simeq 8\pi G a^2 \rho\, ;\label{isorhoembed}\\
    &\frac{2\phi'}{r}\ln a + \left(1-2\phi \right)\left(\frac{\dot a^2}{a^2}-2\frac{\ddot a}{a}\right) \simeq 8\pi G a^2 p_\parallel\, .\label{isopparembed}
\end{align}
\end{subequations}
We now plug these into the conservation equation \eqref{press1}, which, in our fully isotropic case, gives
\begin{equation}
    \frac{\phi''}{r\phi'}-\frac{1}{r^2} =-\frac{1}{\ln a} \left(\frac{\dot a^2}{a^2}+\frac{\ddot a}{a}\right) \, ,
\end{equation}
where we have neglected the higher-order terms and we have assumed $\phi \neq \text{constant}$. The above is characterized by a complete separation between the $r$- and $\eta$-dependent terms. Generally, such equation would have to be solved by setting both terms equal to the same constant. However, this would determine the form of the scale factor. Since we are working at fixed, but fully arbitrary, FLRW cosmological background, the only possibility for the above equation to be consistent is that $\phi = \text{constant}$, which however corresponds to the Minkowski solution. 
This shows that a gravitational system, described by the usual Newtonian weak-field limit with a fully isotropic source, does not allow for a cosmological embedding in FLRW cosmologies, unless the spherical object induces a local anisotropy in the cosmological fluid. 

The derivation above only holds for fully isotropic, Newtonian systems. In particular, it does not apply to Poissonian or non-Poissonian objects. Indeed, in these two cases, \cref{pconserv} does not give an independent equation for $\phi$, $\psi$, $\rho$, but it must be considered as an additional equation defining the tangential pressure $p_{\perp}$. Thus, generally, non-Newtonian systems in the weak-field regime will allow for a cosmological embedding. 

Notice that the above statement remains true for any weak-field non-Newtonian system, even if we impose for the local system an equation of state $p_\parallel=p_\parallel(\rho(r))$ or a given profile for $p_{\perp}(r)$ (this is the case of galaxies, which will be discussed in the next section). 
In fact, in this case, $\phi$ and $\psi$ are independent quantities and \cref{hh1,hh2,pconserv} simply define the (time-dependent) quantities $\rho(\eta,r)$, $p_\parallel(\eta,r)$ and $p_{\perp}(\eta,r)$, respectively.

In order to elucidate these points, let us apply the same method used for the Newtonian system to the non-Newtonian weak-field solution given by \cref{phic1neq0,psic1neq0,gh,pperpNewtoniansolution}.
Using \cref{phic1neq0,psic1neq0} we get  from \cref{pconserv},
\begin{equation}
    -\frac{c_1}{r^2}a^2 - 2r \left[-1 + \ln \left(\frac{r}{\delta} \right) \right]\dot a^2 + r a \left[-1+4\ln \left(\frac{r}{\delta} \right) \right]\ddot a = 8\pi G a^4 \, r\,  p_{\perp} \, .
\end{equation}
This equation has always a solution which determines $p_{\perp}$. 

\subsubsection{Anisotropic Newtonian systems}
\label{subsubsec:AlmostIsoNewt}

In the section above, we have considered the two cases of a fully isotropic, Newtonian system and that of a non-Newtonian, anisotropic system. However, one can also consider local systems in the weak-field regime characterized by an isotropic density profile and small 
anisotropies in the pressures. From a physical point of view, these small anisotropies could naturally be induced by the backreaction between the spherical body and the nearby cosmological fluid.
In order to describe the cosmological coupling of these local systems, we will decompose the density and the pressure profiles of the source as follows (see also Ref.~\cite{Cadoni:2020jxe})
\begin{subequations}
\begin{align}
    \rho(r, \, \eta) & =\frac{\rho(r)}{a(\eta)^2} + \rho_{\mathrm co}(\eta) + \rho^{+}(r, \eta) \, ;\\ 
    p_\parallel(r, \, \eta) & =\frac{p_\parallel(r)}{a(\eta)^2} + p_{\mathrm co}(\eta) + p_{\parallel}^+(r, \eta)\, \label{pTotalParallel};\\
    p_\perp(r, \, \eta) & = \frac{p_\perp(r)}{a(\eta)^2} + p_{\mathrm co}(\eta) + p_{\perp}^+(r, \eta)\, . \label{pTotalPerp}
\end{align}
\end{subequations}

In the above, $\rho(r)$, $p_\parallel(r)$ and $p_\perp(r)$ are given by the field equations in the static and weak-field limits, namely by \cref{rhoWFL,pparWFL,PoissonEqmod}. $\rho_{\mathrm co}$ and $p_{\mathrm co}$ represent, instead, the purely cosmological background contributions, satisfying the Friedmann equations 
\begin{subequations}
\begin{align}
    a^2 \rho_{\mathrm co} &= \frac{3}{8\pi G}\left(\frac{\dot a}{a} \right)^2\, ; \label{rhocosmoiso}\\
    a^2 p_{\mathrm co} & = \frac{1}{8\pi G}\left[\left(\frac{\dot a}{a} \right)^2 - 2\frac{\ddot a}{a} \right]\, .\label{pcosmoiso}
\end{align}
\end{subequations}
Finally, the cross contributions (denoted with the superscript $+$) are the ones which allow for small anisotropies induced on the object when $p_{\parallel}^+\neq p_{\perp}^+ $.

The existence of the cosmologically embedded solutions can be shown by deriving the explicit form of $\rho^{+}$, $p_{\parallel}^+$,  $p_{\perp}^+$. Assuming, for simplicity, the cosmological fluid to be dust, i.e., $p_{\mathrm co} = 0$, from \cref{alpharr}, together with \cref{pcosmoiso,pparWFL}, we obtain
\begin{equation}
p_{\parallel}^+ (r, \eta) = \frac{\phi' }{4 \pi G r a^2} \ln \left(\frac{a}{a_0}\right) \, . \label{eq:pPar+Rel}
\end{equation}
From \cref{alpha00}, instead, we have
\begin{equation}
\rho^+(r,\eta) = -\frac{\nabla^2 \phi}{4 \pi G  a^2} \ln \left(\frac{a}{a_0}\right)  -  \frac{2}{3}\frac{\rho_{\mathrm co}}{r^2} \left(r^3 \phi\right)' \, .
\label{eq:rho+Rel}
\end{equation}
The energy-momentum conservation, to lowest order, gives
\begin{equation}
  p_\perp^+(r, \eta) = \frac{1}{8 \pi G r a^2}\left\{ \left[1 + \ln \left(\frac{a}{a_0} \right) \right](\phi' + r \phi'') + \psi'\right\} + \frac{1}{2} r \phi' \rho_{\mathrm co}\, .
\label{pperplus}
\end{equation}
If the static, local object has isotropic pressure, i.e., $p_\parallel(r) = p_\perp(r) \equiv p(r)$, from \cref{press1}, it can only be stable if $p(r)$ is much smaller than $\rho(r)$, such that $p(r) \sim \alpha' \rho$ (thus implying the Newtonian limit). Following the prescription above, such object can only be cosmologically embedded and continues to be stable (in the sense that $p(r)$ and $\rho(r)$ are kept time-independent quantities) as long as we assume the presence of small, dynamical anisotropies in the pressure, i.e., if $p_\parallel^+ (r, \eta) \not = p^+_\perp(r, \eta)$. This result is perfectly consistent with the absence of a cosmological embedding of a purely-isotropic source at all scales, outlined in the previous subsection. 

Instead, if the static central object is characterized by different pressure components, both of which are subleading with respect to the corresponding density, we recover again \cref{psisolc10}. In this limit, \cref{eq:rho+Rel} can be expressed as
\begin{equation}
\rho^+(r,\eta) = \frac{\rho(r)}{ a^2(\eta)} \ln \left(\frac{a_0}{a(\eta)}\right)  -  \frac{2}{3}\frac{\rho_{\mathrm co}(\eta)}{r^2} \left(r^3 \phi \right)' \, .
\end{equation}
Assuming that $\rho_{\mathrm co}$ is negligible within the spherical body, $a_0$ fixes the instant in which the additional density correction is zero ($\rho^+ = 0$). As $a(\eta)$ increases, $\rho^+$ always decreases and can become negative. If the $\rho_{\mathrm co}$-term is relevant in the weak-field limit, $(r^3 \phi)'$ is expected to be always negative (since $\phi$ would be negative and would approach zero slower than $1/r$). Since $\rho_{\mathrm co } \propto a^{-3}$, at large redshift this term dominates. Therefore, qualitatively, the energy density of such object is expected to increase up to some value of $a$, and then to start decreasing. The turning point should depend on the cosmological background and the object internal structure. This is in qualitative agreement with our general results for the behaviour of the mass $M(a)$ obtained in \cref{subsec:behaviormassformula}. It will be further confirmed in \cref{subsec:rsquaredrhoprofile}.  \\

Summing up, in the weak-field limit, a spherical object with fully isotropic pressure ($p_\perp(r,\eta) = p_\parallel(r),\eta$), immersed in an isotropic fluid, cannot be cosmologically coupled. However, the presence of the spherical body can induce anisotropies in the pressures, i.e., $p^+_\parallel(r, \eta) \not= p^+_\perp(r, \eta)$. We have explicitly shown that non-negligible anisotropies are a necessary ingredient to describe, in a manner consistent with GR, the embedding of local objects into the cosmological background. At the moment, however, we are unable to describe the dynamical mechanism leading to such configurations. If this is the case, there exists a non-trivial cosmological coupling and the MS mass of local objects will evolve cosmologically.

\section{Cosmological coupling of galaxies and galaxy clusters}
\label{sect:galaxies}

Galaxies are understood as bound gravitational systems in the weak-field regime. Here, we consider that the dark-matter halo can be approximated by a spherical matter distribution, as in the standard picture, but which has a non-negligible relativistic pressure. In the standard picture, dark matter is made of particles that only interact through gravity (self-interaction being negligible). In the latter case, the effective pressure of the fluid comes directly from the velocity dispersion of the particles. It is impossible for such non-interacting particles to describe a fluid with relativistic pressure since this would imply relativistic particle velocities, which would far exceed the escape velocity of the galaxy. Self-interacting dark matter does not have this limit.

An important application of our results concerns galaxies (elliptical galaxies in particular) and galaxy clusters. Indeed, these are local objects in the weak-field regime and can be described, in a first approximation, as static and spherically-symmetric. Moreover, their rotational curves and the local velocity dispersion indicate that they can be described as gravitational systems sourced by an anisotropic fluid \cite{Cadoni:2017evg,Tuveri:2019uej,Tuveri:2019zor,Cadoni:2020izk,Cadoni:2021zsl}.
Therefore, in a GR framework, the local dynamics of galaxy and clusters can be described by \cref{alpha00static,alpharrstatic,press1}. In all generality, we expect that they couple with the large-scale cosmological dynamics along the lines discussed in the previous sections.

To close the system of equations \eqref{alpha00static}, \eqref{alpharrstatic}, \eqref{press1}, we need information about their density $\rho(r)$ and pressure ($p_\perp$, $p_{\parallel}$) profiles. Several density profiles for galaxies, devised to reproduce the observed rotational curves, dark-matter halos and/or coming from $N-$body simulations, have been proposed in the literature. Here, we will consider three cases: the simple $1/r^2$, the Navarro-Frenk-White (NFW) and the Einasto profiles.

Providing a description of pressure profiles is much more complicated, as it is very difficult to relate them to direct observations. Galactic pressure profiles are phenomenologically related to the velocity dispersion in the galaxy. A way to codify this information is either through $(1)$ an equation of state $p_{\parallel}=p_{\parallel}(\rho)$ or $(2)$ an explicit profile $p_\perp(r)$.
The simplest assumption one can start from is that tangential pressures can be neglected:
\begin{equation}\label{pres}
p_\perp=0\, .
\end{equation}
Within the standard approach to cold dark matter, the velocity dispersion of the latter is expected to be anisotropic. Hence, this can be seen as a first approximation \cite{1993ESOC...45...79V,Fukushige:2003xc,Gebhardt:2000fk,Hernquist:1990be,Tremaine:2002js}. If one considers a two-fluid model for the galaxy, describing baryonic and dark matter separately, this choice is also quite natural (see \cite{Cadoni:2021zsl} and the discussion therein). By fixing $p_\perp = 0$, we are also fixing the profile of $p_\parallel$ in the weak-field limit. Indeed, \cref{consWFL} implies that $p_\parallel \propto 1/r^2$.

For a given density profile $\rho(r)$, the time-independent metric functions $e^\alpha$, $e^{\beta_0}$ can be derived using \cref{beta0solutiongeneral,SystemEqFirstRegime} and imposing $p_\perp=0$. While $\beta_0(r)$ can be easily obtained by direct integration of \cref{beta0solutiongeneral}, $\alpha(r)$, instead, is obtained by adding together \cref{alpha00static,alpharrstatic}, by differentiating \cref{alpharrstatic} and finally by substituting the results into \cref{press1}. One gets a Riccati equation for $Z\equiv r\alpha'$, namely
\begin{equation}\label{riccati}
Z'+\frac{Z^2}{2r}- \frac{Z\beta_0'}{2}=\beta_0' \, .
\end{equation}
Notice that the function $Z(r)$ coincides with the function $k(r)$ \eqref{kr}, which in turn enters the cosmological mass shift formula \eqref{massshift}.

One can also transform the Riccati equation into a linear second order ODE by defining $ Z=2r(u'/u)$
\begin{equation}\label{secorder}
u'' + \frac{u'}{r}-\frac{\beta_0'}{2}u' -\frac{\beta_0'}{2r}u=0\, .
\end{equation}
Typically, \cref{riccati} or \cref{secorder} cannot be solved in closed form; however, they can be solved perturbatively or numerically. 
Let us discuss separately the $1/r^2$, NFW and Einasto profiles. First, it is instructive to consider the case of the weak-field limit with a constant equation-of-state parameter.

\subsection{Weak-field limit with constant equation-of-state parameter}

As a first application of the derived cosmological coupling with $p_\perp = 0$, we consider the following equation of state  
\begin{equation}
  p_\parallel = w \rho \, ,
\end{equation}
with $w$ a constant. From \cref{consWFL}, which assumes the weak-field limit, one finds $p_\parallel \propto 1/r^2$. Hence, we write the dark-matter density as
\begin{equation}\label{dprof}
\rho(r)= \frac{\rho_0}{r^2}\, .
\end{equation}
This profile was previously considered in Ref.~\cite{Cadoni:2021zsl}. It also describes the large radii behavior of the pseudo-isothermal profile \cite{1989A&A...223...47B}. 

\begin{figure}[!ht]
\centering
\includegraphics[width= 10 cm, keepaspectratio]{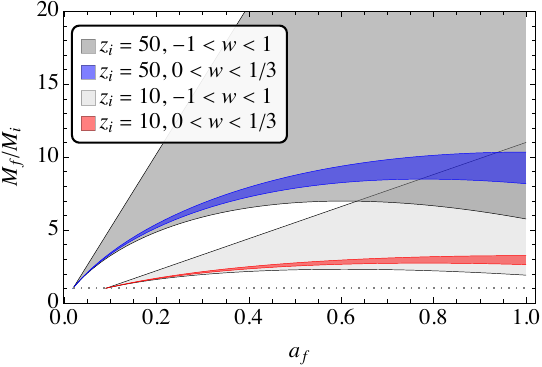}
\caption{The MS mass ratio ($M_f/M_i$) as a function of $a_f$, as given by \cref{eq:MfMiWeakField}. Two different initial redshift values ($z_i$) are considered (10 or 50). For each $z_i$ case, two regions are shown, one in gray for  $-1< w<1$ and the other in blue ($z_i = 50$) or red ($z_i = 10$) for the restricted range $0<w<1/3$. The dotted line is the line with $M_f/M_i = 1$.}
\label{fig:MfMiWeakField}
\end{figure}

From the field equation \eqref{alpharrstatic}, in the weak-field limit we get,
\begin{equation}
    \beta_0(r)  = r \alpha'(r) - 8 \pi G w \rho_0 \, .
\end{equation}

Substituting this solution into the time-time field equation \eqref{alpha00static}  we have,
\begin{equation} \label{weakFieldAlpha}
  \nabla^2 \alpha = 8 \pi G (1 + w) \rho(r) \, .
\end{equation}
For $w \approx 0$, $2 \alpha$ is the Newtonian potential and one recovers the Newtonian limit. The case $w = 0$ implies $p_\parallel = p_\perp = 0$, which is not possible since the spherical object would not be stable. For $w > 0$, it is in general possible to find stability with $p_\parallel \not= p_\perp$, which allows for cosmological coupling.

The $\alpha(r)$ solution, from \cref{weakFieldAlpha} reads,
\begin{equation}
    \alpha(r) = 8 \pi G (1+ w) \rho_0 \ln\frac{r}{r_0} \, ,
\end{equation}
where $r_0$ is an arbitrary constant. In the weak-field limit, as expected, the time-time metric component can be redefined by an additive constant without changing the physical results.

With the $\alpha(r)$ solution, it is possible to explicitly compute $k_L$ and $\beta_0(L)$, which read
\begin{align}
    & k_L   = L \alpha'(L) = 8 \pi G (1+ w) \rho_0 \nonumber \, ,\\
    & \beta_0(L)  = L \alpha'(L) - 8 \pi G w \rho_0 =  8 \pi G \rho_0 \, .
\end{align}
We point out that both $\beta_0$ and $k_L$ do not depend on $L$ in this case. Thus, the MS mass reads
\begin{align}
  M(\eta_f,L)  &= M(\eta_i,L)\frac{a_f}{a_i} \frac{ 1-e^{-\beta_0(L) + k_L \ln a_f}}{1-e^{-\beta_0(L) + k_L \ln a_i}}  \nonumber \\[.2cm]
 & = M(\eta_i,L)\frac{a_f}{a_i} \frac{1 - e^{8 \pi G \rho_0(-1 + (1+w)\ln a_f)}}{1 - e^{8 \pi G \rho_0(-1 + (1+w)\ln a_i)}} \, \label{eq:MfMiWeakField}\\[.2cm]
  & \approx M(\eta_i,L)\frac{a_f}{a_i} \frac{1 - (1+w)\ln a_f}{1 - (1+w)\ln a_i} \nonumber \, . 
\end{align}

The above expression is meaningful only when there is an adequate pressure value to support the spherical object, so that the case $ w = 0$, as an exact expression, is not a valid case; however, small and positive $w$ values are possible. We add that there are exotic compact objects with $w \approx -1$ that can be stable and, in this case, \cref{eq:MfMiWeakField} implies a linear increase of the mass. \Cref{fig:MfMiWeakField} shows the evolution of $M_f/M_i$ with different conditions. 

\subsection{$1/r^2$ density profile}
\label{subsec:rsquaredrhoprofile}

In this section, we consider three cases related 
to this halo profile. First we consider the weak-field limit, then we consider two classes of solutions that go beyond the weak-field limit.

\subsubsection{Exact case}

This profile is not able to correctly describe the observed phenomenology near the galactic center, since $\rho$ diverges as $1/r^2$ for $r\to 0$. Therefore, we will employ it here given its simplicity, as it allows us to analytically solve our equations. This will enable us to test the cosmological coupling of non-Newtonian systems in the weak-field limit in a controlled manner. 

Instead of solving the Riccati equation \eqref{riccati} or the associated second-order equation \eqref{secorder}, in this case $\alpha$ can be found by recasting \cref{alpharrstatic,press1} into the following forms 
 \begin{equation}
\alpha'= -  \frac{2}{r^2}\frac{\frac{\dd}{\dd r} \left(r^2p_{\parallel}\right)} {\rho+p_{\parallel}},\quad - \frac{1}{r^2} \frac{\dd}{\dd r} \left(r^2p_{\parallel}\right)(r-2GM)=
\left(8\pi G r^2 p_{\parallel} + \frac{2GM}{r}\right)(\rho+p_{\parallel})\, .
\label{static1}
\end{equation}
We can now use the density profile \eqref{dprof} into \cref{beta0solutiongeneral} to get the mass function $m(r)$ and $\beta_0$
\begin{equation}
m(r)= 4\pi\sigma r + m_B\, , \qquad e^{-\beta_0(r)} = 1-\frac{2Gm_B}{r}-8\pi G \sigma\,.
\label{static}
\end{equation}
$m_B$ is an integration constant that might be interpreted as the baryonic matter. This contribution can be neglected as long as we consider large galactic scales over $\sim 20 \ \text{kpc}$ (where rotation curves starts flattening, according to observations). 

A quick inspection of the right equation in \cref{static1} reveals that there are two families of solutions. 

\subsubsection{First family: $p_\parallel = -\rho$}

In this case, $p_\parallel= -\rho=-\sigma/r^2$ automatically satisfies the $p_\parallel$ equation in \cref{static1}. Therefore, plugging this result into \cref{alpharr} gives
\begin{equation}\label{sol2}
r\alpha'(1-8\pi G \sigma)= 0 \, \qquad \Rightarrow \qquad \alpha = \text{constant}\, , 
\end{equation}
where the constant can be set equal to zero without loss of generality (we can always perform a time rescaling to absorb it).

\subsubsection{Second family}
The general solution of \cref{static1} for $p_\parallel\neq -\rho$ together with \eqref{static} reads as
\begin{equation}\label{alphasecondsolution}
p_\parallel=-\frac{\sigma}{r^2}+\frac{1-8\pi G \sigma}{4\pi G r^2 \ln\left(r/\ell \right)}\, , \qquad  e^{\alpha} = B \, \ln\left(\frac{r}{\ell} \right)\, ,
\end{equation} 
where $\ell$, $A$ are integration constants. 
Additionally, in order for $e^\alpha$ to be positive, we must require $r > \ell$, which corresponds to the regime of validity of this solution (for example, this can be phenomenologically identified as the regime in which the baryonic matter is negligible and the rotational curves begin to flatten out). 

We see that the solutions of the first family, characterized by $p_{\parallel}=-\rho$ and a negative pressure, dominate for $r\to \infty$, i.e., in the transition to the cosmological regime. Conversely, at smaller (galactic) scales, the second positive term for the pressure dominates. Physically, this means that, at galactic scales, the hydrostatic equilibrium is reached when the dark-matter halo balances the gravitational attraction through positive radial pressure. On the other hand, at great distances, at the transition to the cosmological regime, the hydrostatic equilibrium is not reached in the way we are used to. It is a local equilibrium in which both sides of the hydrostatic equilibrium equation (right equation in \cref{static1}) vanish separately. 

In the generic case, our static solution describes a spacetime with a conical singularity \cite{Cadoni:2021zsl}. In the simplest case given in \cref{sol2}, the metric is 
\begin{equation}
\dd s=-\dd\eta^2+ \frac{\dd r^2}{(1-8\pi G \sigma)}+r^2\dd\Omega^2\, .  
\label{metric}
\end{equation}
The solution is physically acceptable in the weak-field limit when the deficit angle is very small
  \begin{equation}
 \sigma\ll\frac{1}{8\pi G}. 
\label{wf}
\end{equation} 
In this limit, the spacetime can be well approximated by the Minkowski one.
 
By virializing the galactic motion, one gets the asymptotic constant velocity $ v^2= 2 G\sigma$, and $\sigma\approx r a_0/G$, where $a_0$ is the typical acceleration scale \cite{Milgrom:1983ca,Milgrom:1983pn,deMartino:2020gfi,McGaugh:2020ppt}. From observations we know that $a_0\approx H=\frac{1}{\hat L}$, where $\hat L$ is the cosmological horizon. From condition \eqref{wf}, it follows that $r \ll \hat L$, which is always satisfied at galactic scales.

\subsubsection{Cosmological coupling of galaxies with $1/r^2$ density profile}

\begin{figure}[!t]
\centering
\includegraphics[width= 13 cm, height = 13 cm,keepaspectratio]{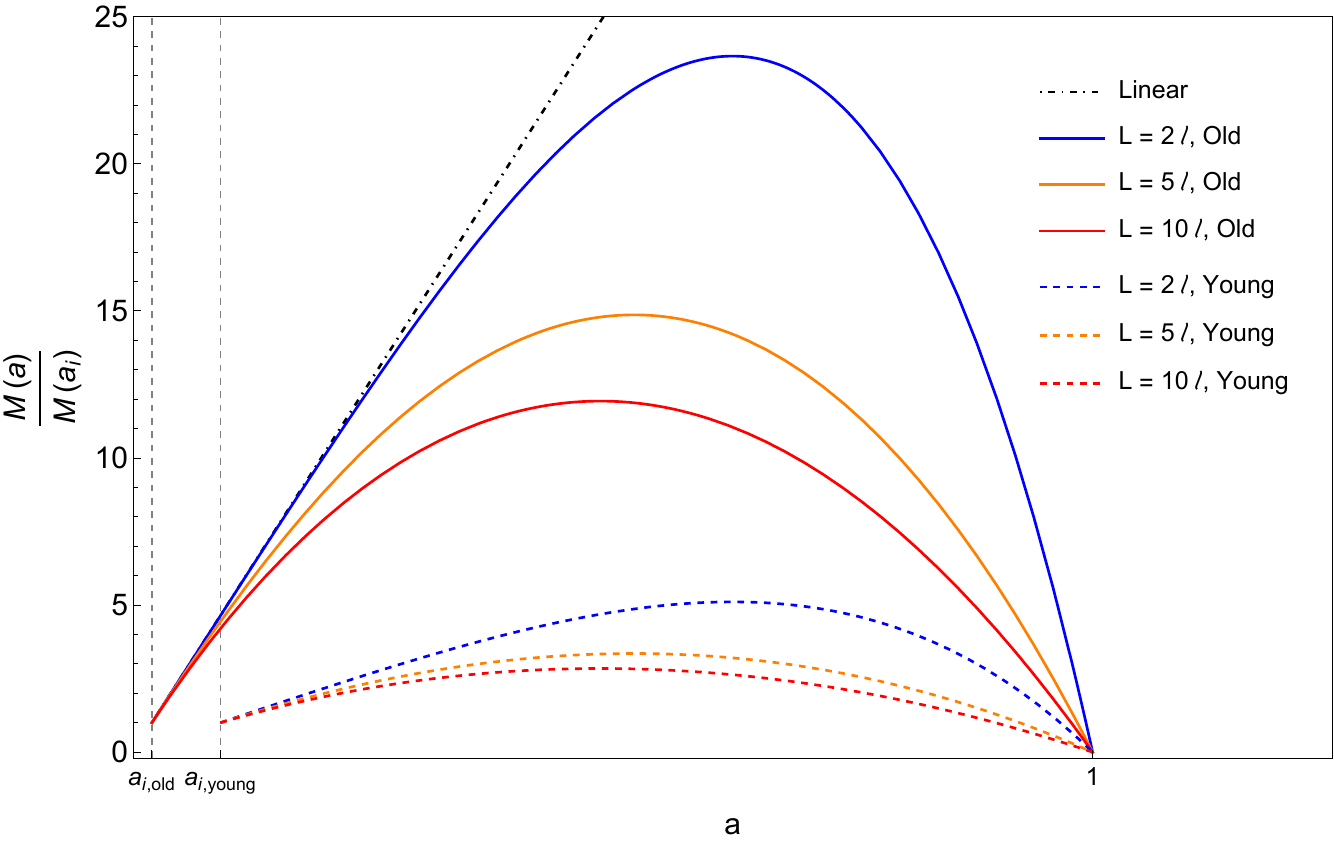}
\caption{Plot of \cref{massshiftrhor2} as a function of the scale factor $a$, for different values of $L/\ell$. The solid curves are obtained by evaluating \cref{massshiftrhor2} at redshift $z = 50$, representing the epoch of formation of old galaxies. The dashed curves are obtained by evaluating \cref{massshiftrhor2} at $z = 10$, representing the epoch of formation of young galaxies. We also plot for comparison the linear scaling $M(a) \propto a$ (dash dotted black line). We set $8\pi G \sigma \sim 10^{-6}$.}
\label{fig:Fzsigma1r2}
\end{figure}

From the results of \cref{sect:cc,sect:ce}, we can discuss the cosmological coupling of galaxies and evaluate the mass shift \eqref{massshift} for the two families of solutions found above.

\subsubsection*{First family}

In this case, working in the weak-field limit for which \cref{wf} holds, we can neglect the deficit angle. Thus, the solution reduces essentially to the Minkowski space and the density and pressures are zero. The weak-field limit is Newtonian, so there is no cosmological coupling of the solution. 

\subsubsection*{Second family}
In this case, the pressure $p_\parallel(r)$ and the density are not zero, the weak-field limit is non-Newtonian, hence the cosmological embedding of the solution is possible.
From \cref{alphasecondsolution}, we have
\begin{equation}
k(r) = r\alpha'(r) = \frac{2}{\ln\left(r/\ell \right)}\, ,
\end{equation}
which should be evaluated at a given galactic radius $r = L$, representing the size of the galaxy. Regardless of the value of $L$, we see that $k(r) > 0$ since $r>\ell$ (see \cref{alphasecondsolution}). Therefore, the cosmological mass shift \eqref{massshift} reads 
\begin{equation}\begin{split}\label{massshiftrhor2}
\frac{M(a)}{M(a_i)} = \frac{a}{a_i} \frac{1-(1-8\pi G \sigma)a^{k_{L}}}{1-(1-8\pi G \sigma)a_i^{k_{L}}}\, .
\end{split}\end{equation}
$8\pi G\sigma$ is a small parameter but must be kept different from zero. It acts as a kind of regulator for the the mass $M(a)$, as it prevents the latter from going to zero at the present time, $a = 1$. Additionally, it allows to recover the linear scaling $M(a) \propto a$ in the limit $k_L \to 0$. 

Since $e^{-\beta_0}$ is very close to $1$, we expect the inequality \eqref{ineq} to hold. Indeed, there is a maximum at 
\begin{equation}
 a_M = [1+k_L - 8\pi G \sigma (1+k_L)]^{-1/k_L}\simeq (1+k_L)^{-1/k_L}\, ,
\end{equation}
where we have neglected the deficit angle. By varying the values of the galactic radius $L$ and selecting (small) values of the parameter $8\pi G\sigma$, we show some curves for $M(a)$ in \cref{fig:Fzsigma1r2}. 

$M(a)$ behaves in an unphysical way. The mass grows linearly near its formation and keeps growing in a sub-linear way until it reaches a maximum; then it starts decreasing.   
This behavior is a consequence of the validity of the inequality \eqref{ineq} and gives a further indication that such a simple $r^{-2}$ density profile is not appropriate for a physically realistic description of galactic dynamics.
 
\subsection{Navarro-Frenk-White density profile}
The NFW density profile is employed to fit dark-matter profiles derived from $N-$body simulations. It gives a more realistic description of galactic dark-matter halos than the rough $r^{-2}$ profile. It reads as
\begin{equation}\label{dprof1}
\rho(r)= \frac{\rho_c}{\frac{r}{r_c}\left(1+\frac{r}{r_c}\right)^2}\, ,
\end{equation}
where $\rho_c$ and $r_c$ are two parameters characterizing the model.

Although the $1/r$ behavior of the NFW density profile near the galactic center represents an improvement with respect to the $1/r^2$-divergence of the model analyzed in the previous section, it still does not provide a satisfactory description of halos in the bulge region (see Ref.~\cite{Salucci:2018hqu}).

$r_c$ is the typical length-scale at which dark-matter effects begin to become relevant, whereas $\rho _c$ is related to the critical density of the universe $\rho_\text{critic} = 3H^2/8\pi G$, where $H$ is the Hubble constant. Both $\rho_c$ and $r_c$ are defined in terms of a reference scale, which is the virial radius, commonly known as $r_{200}$. It physically represents the radius within which the halo density corresponds approximately to $200$ times the critical density of the universe. $r_{200}$ is inferred from the mass of the halo $M_{200} = 200 \rho_\text{critical} (4\pi/3)r_{200}^3$.

The ratio of $r_{200}$ and $r_c$ also defines a relevant parameter $\mathcal{C} \equiv r_{200}/r_c$, called the concentration, which can actually be measured or inferred from numerical simulations. Considering another parameter $\delta_c$ (the overdensity parameter), the NFW profile can be rewritten as follows
\begin{equation}
\frac{\rho(r)}{\rho_\text{critical}} = \frac{\delta_c}{\frac{r}{r_c}\left(1+\frac{r}{r_c} \right)^2}\, .
\end{equation}
By requiring that, at $r_{200}$, $\rho = 200 \rho_\text{critical}$, we also obtain the expression of $\delta_c$
\begin{equation}
\delta_c = \frac{200}{3}\frac{\mathcal{C}^3}{\log\left(1+\mathcal{C}\right)-\frac{\mathcal{C}}{1+\mathcal{C}}}\, ,
\end{equation}
which also defines $\rho_c$ as 
\begin{equation}
\rho_c = \delta_c \, \rho_\text{critical}\, .
\end{equation}
Summarizing, the quantities measured/inferred from numerical simulations are $\rho_\text{critical}$, $r_{200}$, $\delta_c$  and $\mathcal{C}$. In terms of them, we can express the parameters characterizing the NFW profile \eqref{dprof1}:
\begin{equation}
r_c= \frac{r_{200}}{\mathcal{C}} \, , \qquad \rho_c = \delta_c \, \rho_\text{critical}\, .
\end{equation}
Using $r_c$, $\rho_c$, we can construct a dimensionless parameter\footnote{Reinstating the speed of light $c$, it should read as $8\pi G \rho_c r_c^2/c^2$.},  
\begin{equation}\label{kdl}
\mathcal{K} \equiv 8\pi G \rho_c r_c^2\, ,
\end{equation}
which will be used below to describe the cosmological mass shift of galaxies. 

Some values of the parameters  $M_{200}$, $r_{200}$, $\mathcal{C}$, $r_c$, $\delta_c$, $\mathcal{K}$ are given in \cref{tab:NFWparameters} for a sample of galaxies taken from \cite{Navarro:1995iw}. 

\begin{table}[]
    \centering
    \begin{tabular}{llllll}
        \toprule[1pt]
        $M_{200} [10^{12}\, M_\odot]$ \quad &   $r_{200} [\text{kpc}]$ \quad & $\mathcal{C}^{-1} =r_c/r_{200}$ & $r_{c} [\text{kpc}]$ & $\delta_c$ & $\mathcal{K}$\\
        \midrule[0.5pt]
       $0.319$ \quad &   $177$ \quad & $0.052$ & $9.204$ & $230538$ & $3.46\cdot 10^{-7}$\\
        $2.425$ \quad &   $348$ \quad & $0.06$ & $20.88$ & $160060$ & $1.24\cdot 10^{-6}$\\
        $2.552$ \quad &   $354$ \quad & $0.124$ & $43.896$ & $26596.3$ & $9.09 \cdot 10^{-7}$\\
        $29.67$ \quad &   $802$ \quad & $0.131$ & $105.062$ & $23322.9$ & $4.57 \cdot 10^{-6}$\\
   $102.68$ \quad &   $1213$ \quad & $0.110$ & $133.43$ & $35504.7$ & $1.12 \cdot 10^{-5}$\\
   $1109.9$ \quad &   $2682$ \quad & $0.151$ & $404.982$ & $16659.5$ & $4.85\cdot 10^{-5}$\\
   $1931.5$ \quad &   $3226$ \quad & $0.188$ & $606.488$ & $10014.7$ & $6.53 \cdot 10^{-5}$\\
   $3009.7$ \quad &   $3740$ \quad & $0.143$ & $534.82$ & $18940.5$ & $9.61 \cdot 10^{-5}$\\
        \bottomrule[0.8pt]
    \end{tabular}
    \caption{Table of astrophysical parameters related to the NFW profile, according to the data contained in the original paper \cite{Navarro:1995iw}.}
    \label{tab:NFWparameters}
\end{table}

Using the NFW density profile \eqref{dprof1} into \cref{beta0solutiongeneral}, one easily finds the mass function $m(r)$ and the metric function $\beta_0$

\begin{equation}\label{massNFW}
m(r)= \frac{{\cal K} r_c}{2}\left[\ln\left(1+\frac{r}{r_c}\right)-\frac{\frac{r}{r_c}}{1+\frac{r}{r_c}}\right]\, , \qquad e^{-\beta_0}=1- \frac{{\cal K}r_c}{r}\left[\ln\left(1+\frac{r}{r_c}\right)-\frac{\frac{r}{r_c}}{1+\frac{r}{r_c}}\right],
\end{equation}
where ${\cal K}$ is given by \cref{kdl} and we have fixed the integration constant such that $m(0)=0$.

In order to find the second metric function $\alpha$, we have either to solve the Riccati equation \eqref{riccati} or the associated second order equation \eqref{secorder}. 
This cannot be done in closed analytic form for the present case.
However, to compute the cosmological mass shift \eqref{massshift}, we need to evaluate the solution at the galactic radius $r=L$. Being $L\gg r_c$, we just need approximate solutions for $r/r_c\gg 1$. Expanding around $r/r_c\to \infty$ we get
\begin{equation}\label{metricexp}
e^{-\beta_0}=1- {\cal K}\frac{r_c}{r}\ln\frac{r}{r_c} +{\cal O}\left(\frac{r_c}{r}\right)\, .
\end{equation}
Using \cref{metricexp} into the Riccati equation \eqref{riccati}, we obtain the approximate solution for $Z=r\alpha'$
\begin{equation}\label{zapprox}
Z(r)=k(r)={\cal K}\, r_c\, \frac{\ln\frac{r}{r_c}}{r}+{\cal O}\left(\frac{r_c}{r}\right)\, .
\end{equation}
Finally, this gives the exponent $k(L)\equiv k_L$ in this approximation
\begin{equation}\label{kapprox}
k_L= {\cal K} \, \frac{r_c}{L} \ln\frac{L}{r_c} \, .
\end{equation}
\subsubsection{Cosmological mass shift}
The static spherically-symmetric solution given by \cref{massNFW,metricexp,zapprox} has a non-Newtonian weak-field limit. According to the results of \cref{sect:ce}, it, thus, allows for a cosmological embedding in a FLRW cosmology. We thus calculate the cosmological mass shift for the galaxy. By virtue of \cref{metricexp,kapprox,massshift}, it gives
\begin{equation}\label{massshiftNFK}
M(\eta_f,L)= M(\eta_i,L)\frac{a_f}{a_i} \frac{ 1-\left[1-k\right]a_f^{k_L}}{ 1-\left[1-k_L\right]a_i^{k_L}}\, ,
\end{equation}
whereas \cref{GeneralMassFormula} gives 
\begin{equation}\label{mass}
M(a)= \frac{L a}{2G} \left[1-(1-k_L) e^{k_L\ln a}\right]\, ,
\end{equation}
with $k_L$ given by \cref{kapprox}. 

The behavior of the mass can be easily inferred by inserting \cref{metricexp,kapprox} into \cref{ineq}.
One finds that the inequality is always violated, thus the mass increases monotonically with the scale factor.

\begin{figure}[!ht]
\centering
\includegraphics[width= 12.5 cm, height = 12.5 cm,keepaspectratio]{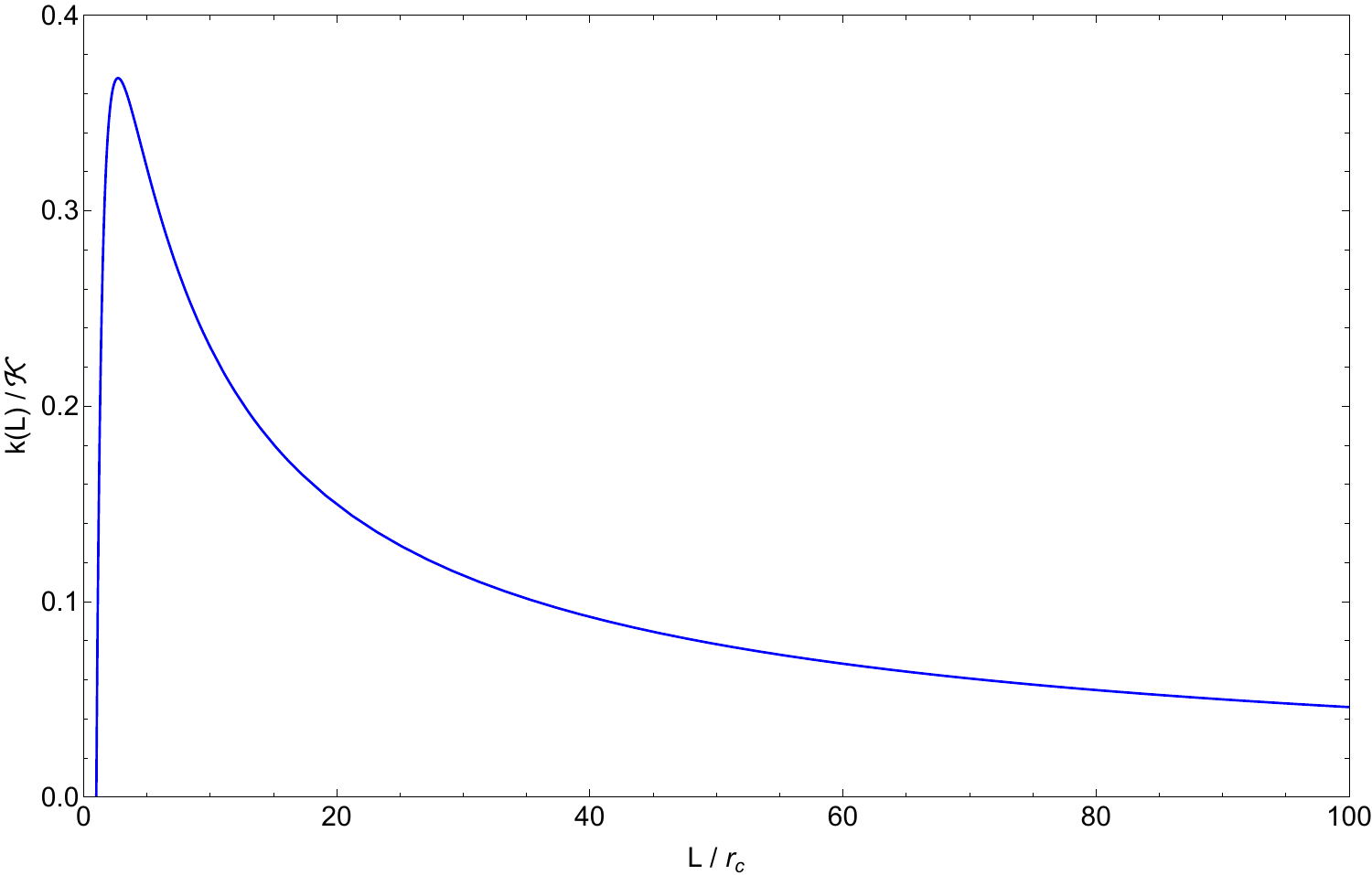}
\caption{Plot of $k_L/\mathcal{K}$ as a function of $L/r_c$ for the NFW profile. }
\label{fig:KfigureNFW}
\end{figure}

Introducing the dimensionless ratio $y=L/r_c$, the parameter $k_L$ given by \cref{kapprox} can also be written as
$k_L = \mathcal{K}(\ln y/y)$. Even if the scale $L$ is not really determined (but presumably it should be of the order of $r_{200}$), we should expect $y\gg 1$ to hold. Therefore, the function $y^{-1} \ln y$ is bounded from above by $\text{e}^{-1} \simeq 0.367879$ since it has a maximum at $\bar y = \text{e}$. We show the plot of $k_L/\mathcal{K}$ in \cref{fig:KfigureNFW}.
In the best case scenario, namely for very massive galaxies (see the last lines in  \cref{tab:NFWparameters}) and $y \sim \text{e}$ (maximum of \cref{fig:KfigureNFW}), we should expect a $k_L$ of the order $10^{-5}$. Actually this estimate is not far from being true, if we identify $L$ with $r_{200}$. Indeed, if we take, e.g., the last line in \cref{tab:NFWparameters}, we have that $y = r_{200}/r_c \sim 6.99301$. In this case we have therefore $k_L \simeq 2.7 \cdot 10^{-5}$.
This also explains why the inequality \eqref{ineq} cannot be satisfied. Indeed, \cref{mass} has a maximum at $a_M = (k_L + k_L^2)^{-1/k_L}$, which for very small $k_L$ goes far beyond $a = 1$.
\begin{figure}[!ht]
\centering
\includegraphics[width= 13 cm, height = 13 cm,keepaspectratio]{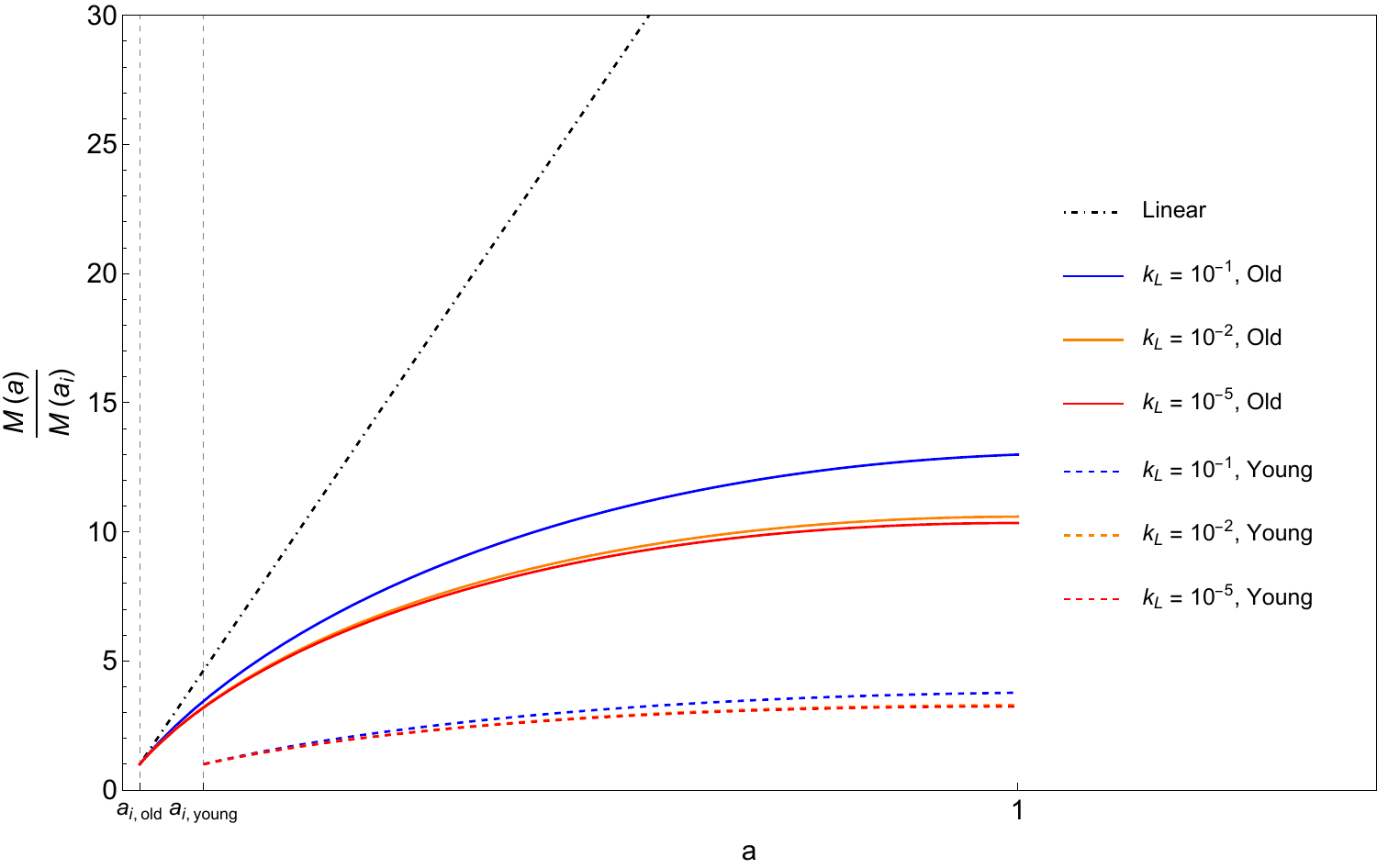}
\caption{Plot of \cref{massshiftNFK} as a function of the scale factor $a$ for different values of $k_L$. The solid curves are obtained by evaluating \cref{massshiftNFK} at redshift $z = 50$, representing the epoch of formation of old galaxies. The dashed curves are obtained by evaluating \cref{massshiftNFK} at $z = 10$, representing the epoch of formation of young galaxies. We also plot for comparison the linear scaling $M(a) \propto a$ (dash dotted black line).
}
\label{fig:MaNFW}
\end{figure}
In \cref{fig:MaNFW} we provide the plots of the function $M(a)$ for different values of the parameter $k_L$. In the physical range $a_i<a<1$, the mass grows monotonically with $a$. The growth starts linearly near $a_i$, but becomes almost immediately strongly sublinear, due to the smallness of $k_L$. The larger values of $k_L$ used in the figure are unrealistic. They were adopted just to get a qualitative idea of the behavior of the mass shift when it becomes relatively large. 

\subsection{Einasto profiles}

Another density profile that can be used to model dark matter halos in galaxies, which can be roughly described as spherically-symmetric, is the Einasto profile
\begin{equation}\label{Einastodensity}
    \rho = \rho_0 \, e^{-\left(r/r_c \right)^{1/n}}\, ,
\end{equation}
where $n$ is a positive index and $r_c$ a characteristic length scale of the profile. $n=1$ is typically used to describe galactic disks, whereas $n=4$ is used for elliptical galaxies. In general, we can consider that the mass $M$ of the halo is contained within a radius $R$, so that the mass function \eqref{beta0solutiongeneral} can be written in the more convenient way
\begin{equation}\label{ggg}
    m(r) = M \frac{\Gamma\left(3n \right)-\Gamma\left[3n, \left(r/r_c \right)^{1/n} \right]}{\Gamma\left(3n \right)-\Gamma\left[3n, \left(R/r_c \right)^{1/n} \right]}\, ,
\end{equation}
where $\Gamma(a,x)$ is the incomplete Gamma function. 

We will use the same strategy for both cases $n=1$ and $n=4$: first solve the Riccati equation \eqref{riccati} approximately near the origin $r \sim 0$, then the numerical result will be used inside \cref{massshift} to infer the behavior of the cosmological mass shift. The first step will then be employed as a boundary condition at the point $r=\epsilon$ to determine the numerical solution of \cref{riccati} up to a certain distance from the center of the halo.
Similarly to the NFW profile, also here the GR solution possesses a non-Newtonian weak-field limit, allowing for its cosmological embedding.
Let us discuss separately the $n=1$ and $n=4$ cases.
\begin{figure}[!ht]
\centering
\includegraphics[width= 13 cm, height = 13 cm,keepaspectratio]{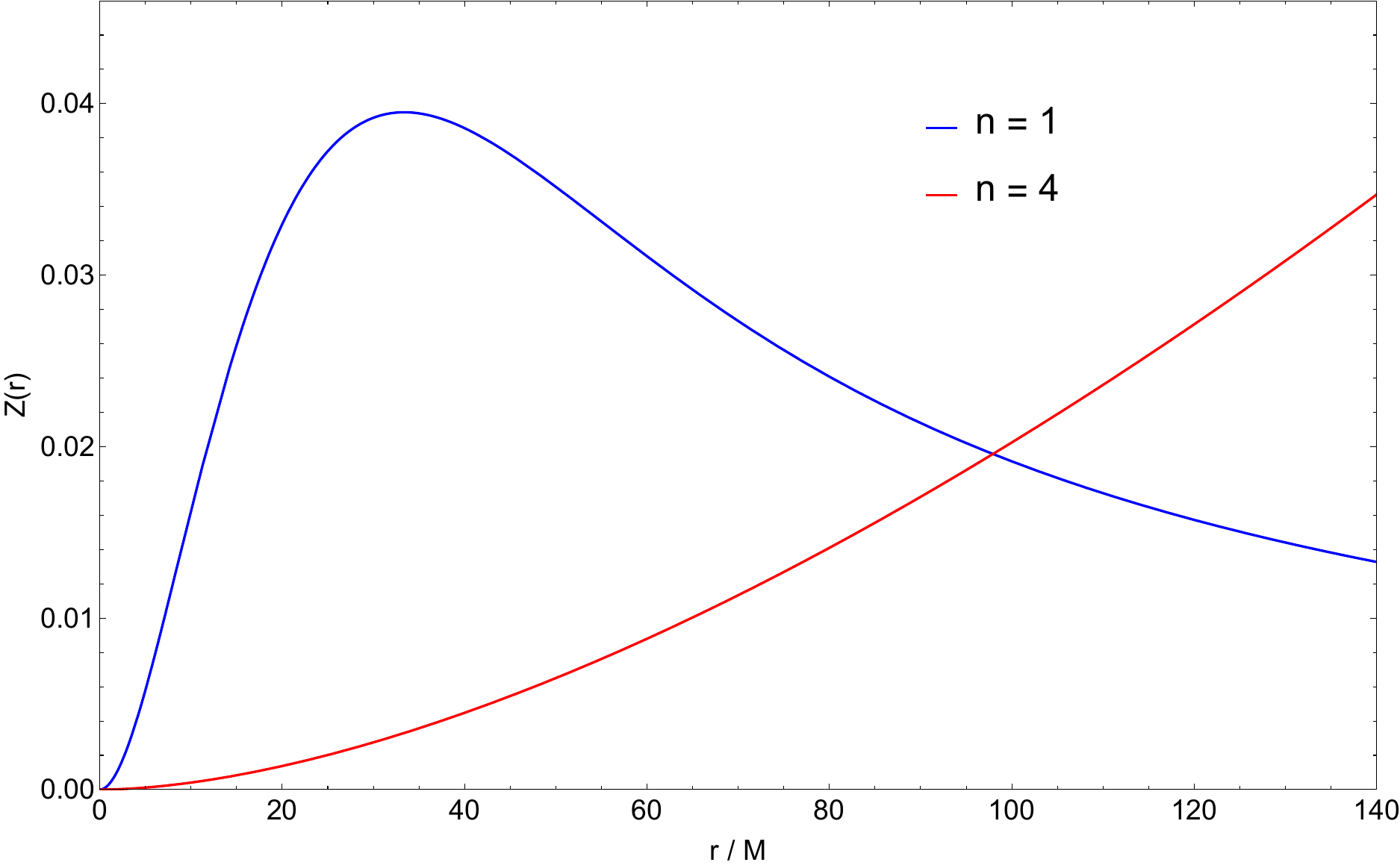}
\caption{Plot of the numerical solution of the Riccati equation \eqref{riccati} with the Einasto profile \eqref{Einastodensity} with indexes $n = 1$ (blue curve) and $n = 4$ (red curve).}
\label{fig:Einaston1Z4Zfigure}
\end{figure}

\subsubsection{$n = 1$}
Using \cref{ggg} into \cref{beta0solutiongeneral} and expanding near the galactic center, i.e., $r=0$, the metric function $\beta$ reads
\begin{equation}\label{jjj}
    e^{-\beta_0} \simeq 1+\frac{2MG r^2}{3r_c^3 \left[-2 + \Gamma\left(3, \frac{R}{r_c} \right) \right]} + \mathcal{O}(r^3) \, , \qquad \beta_0 \simeq -\frac{2G M \, r^2}{3r_c^3 \left[-2 + \Gamma\left(3, \frac{R}{r_c} \right) \right]}\, .
\end{equation}

It is also easy to derive the corresponding approximate solution of the Riccati equation \eqref{riccati}, 
\begin{equation}\label{ZEinastoboundaryn1}
    Z(r) \simeq -\frac{4GM \, r}{3r_c^3 \left[-2 + \Gamma\left(3, \frac{R}{r_c} \right) \right]}\, .
\end{equation}

We have all the ingredients to solve \cref{riccati} numerically. It is of interest to look into the qualitative behavior of the cosmological mass shift \eqref{massshift}. We will consider selected values of the parameters:
\begin{equation}\label{parametersnumerical}
    R = 100 \, M\, ; \qquad r_c = 10 \, M\, ; \qquad M = 1\, ; \qquad \epsilon = 10^{-6}\, . 
\end{equation}
The result of the numerical integration is plotted in \cref{fig:Einaston1Z4Zfigure}.
Notice the similarity of the curve in \cref{fig:Einaston1Z4Zfigure} with the behavior of $k_L(y)$ of the NFW profile.

\begin{figure}[!ht]
\centering
\includegraphics[width= 14 cm, height = 14 cm,keepaspectratio]{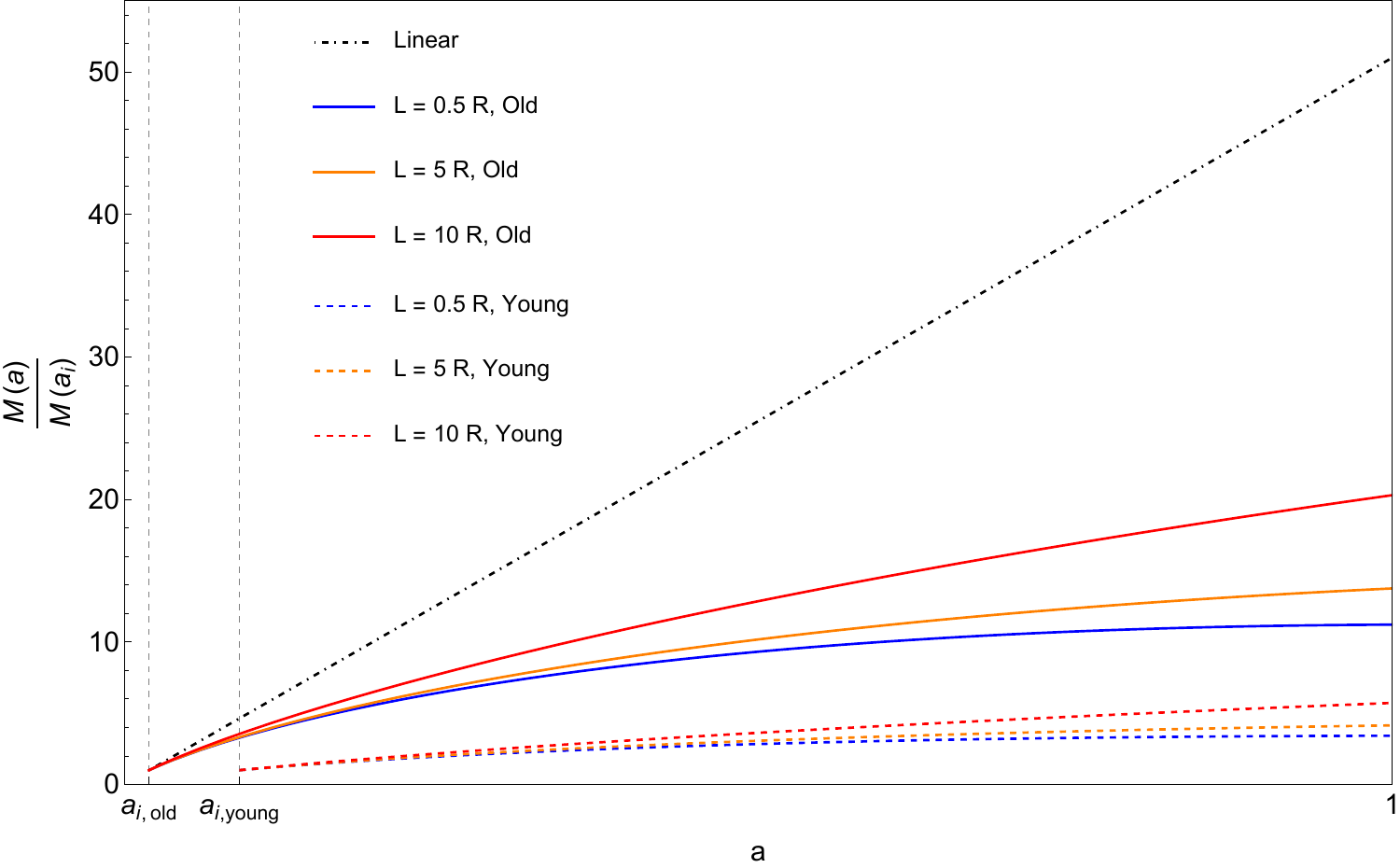}
\caption{Plot of \cref{ggg} as a function of $a$, for the Einasto profile with $n=1$. The solid curves are obtained by evaluating \cref{ggg} at redshift $z = 50$, representing the epoch of formation of old galaxies. The dashed curves are obtained by evaluating \cref{ggg} at $z = 10$, representing the epoch of formation of young galaxies. We also plot for comparison the linear scaling $M(a) \propto a$ (dash dotted black line).} 
\label{fig:MaEinaston1}
\end{figure}

This result is then used to compute the mass shift \eqref{massshift}. The result is plotted in \cref{fig:MaEinaston1}. 

The mass $M(a)$ grows monotonically in the physical range $a_i<a<1$, similarly to what happens for the NFW profile. However, here the curves are not only strongly sublinear, but tend to flatten at values of $a$ close to 1.  

\subsubsection{$n = 4$}

We essentially repeat the same steps used for $n = 1$, taking now $n = 4$. The expansion of the metric function  $\beta$ near $r = 0$ now reads
\begin{equation}
    e^{-\beta_0} \simeq 1+\frac{G M \, r^2}{6 r_c^3 \left[-39916800 + \Gamma\left(12, \frac{R^{1/4}}{r_c^{1/4}} \right) \right]} + \mathcal{O}(r^3) \, , \qquad \beta_0 \simeq -\frac{G M \, r^2}{6 r_c^3 \left[-39916800 + \Gamma\left(12, \frac{R^{1/4}}{r_c^{1/4}} \right) \right]}\, .
\end{equation}

The near-$r=0$ approximate solution  of \cref{riccati} for $Z$ now reads
\begin{equation}
    Z \simeq -\frac{G M \, r^2}{6 r_c^3 \left[-39916800 + \Gamma\left(12, \frac{R^{1/4}}{r_c^{1/4}} \right) \right]}\, .
\end{equation}

For the numerical integration, we adopt the same parameters as in \cref{parametersnumerical}. The result is plotted in \cref{fig:Einaston1Z4Zfigure}. 

\begin{figure}[!ht]
\centering
\includegraphics[width= 14 cm, height = 14 cm,keepaspectratio]{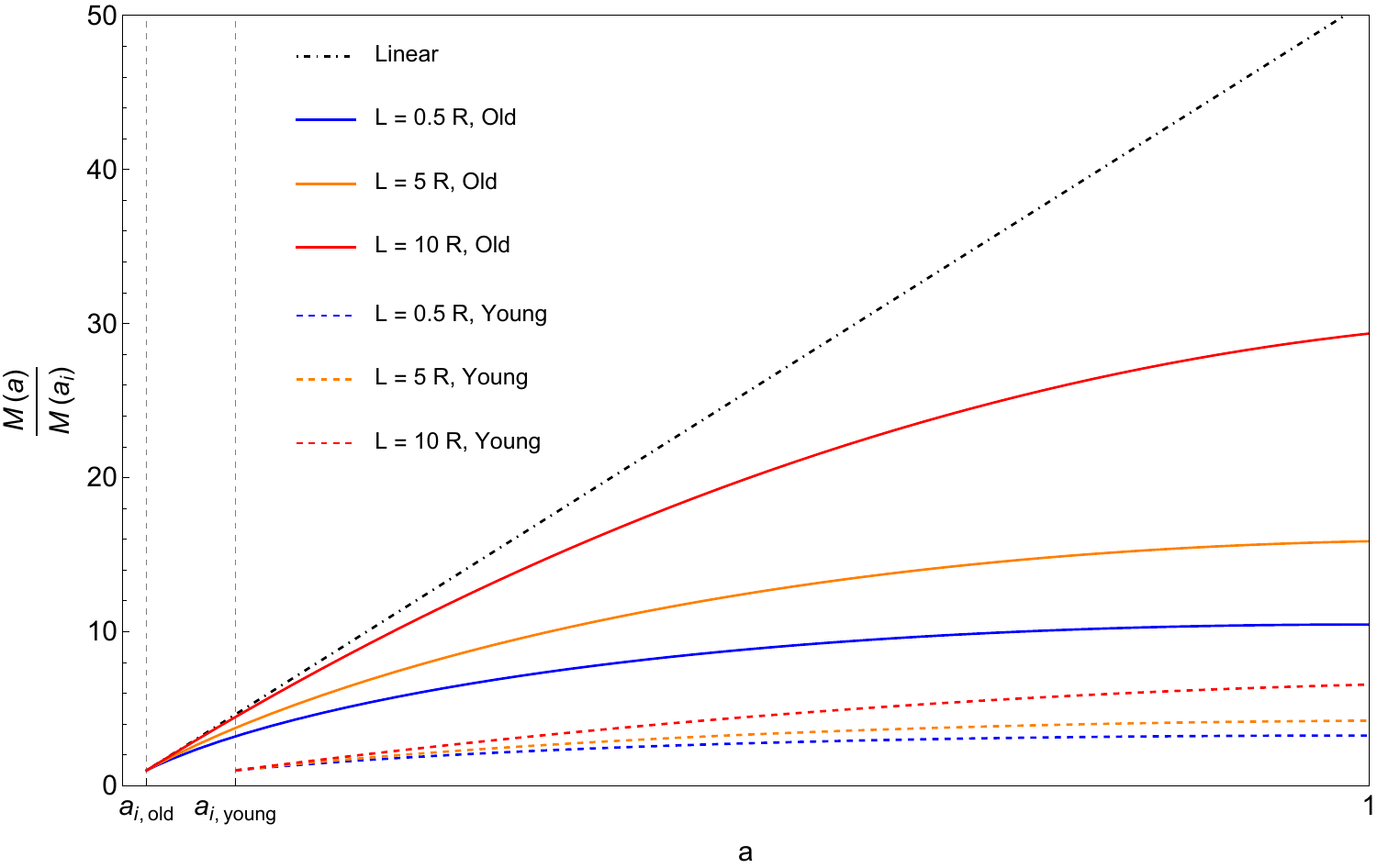}
\caption{Plot of \cref{ggg} as a function of $a$, for the Einasto profile with $n=4$. The solid curves are obtained by evaluating \cref{ggg} at redshift $z = 50$, representing the epoch of formation of old galaxies. The dashed curves are obtained by evaluating \cref{ggg} at $z = 10$, representing the epoch of formation of young galaxies. We also plot for comparison the linear scaling $M(a) \propto a$ (dash dotted black line).}
\label{fig:MaEinaston4}
\end{figure}

Finally, the mass shift \eqref{massshift} is plotted in \cref{fig:MaEinaston4}.
The behavior of $M(a)$ is again quite similar to the NFW and $n=1$ Einasto profiles. It is characterized by a monotonic and strongly sublinear growth and flattening.

\section{Conclusions }
\label{sec:Conclusions}

In this paper we have derived a general formula for the MS mass giving the cosmological coupling of local, static, spherically-symmetric, cosmologically-embedded astrophysical objects. This formula generalizes the one given in Refs.~\cite{Cadoni:2023lum,Cadoni:2023lqe} for black holes or other ultra-compact objects. We have also proven general statements concerning the cosmological embedding of weak-field local systems sourced by anisotropic fluids. We have shown that anisotropies in the pressures, which could be induced by the influence of the cosmological background, are essential to allow for a cosmological coupling. Although adopting  simple and not completely realistic models to describe the halo density profiles, by applying our results to galaxies, we have shown that their mass can be cosmologically coupled. Our results are rather solid and general. Generically, the MS mass formula holds true once the cosmological embedding is possible. The latter follows from a quite general feature, namely the existence of a non-Newtonian weak-field limit.
Moreover, our results are consistent with what is already known about the cosmological coupling of black holes and other strongly-coupled objects, for which the anisotropy in the stress-energy tensor seems to be a necessary condition to allow the cosmological coupling \cite{Cadoni:2023lum,Cadoni:2023lqe}. 

Let us briefly discuss some drawbacks of our approach and related open issues. Apart from black holes, black-hole mimickers and galaxies/galaxy clusters, one would like to test the cosmological coupling also with other local objects in the weak-field regime, like stars with anisotropies (see, e.g., Refs.~\cite{Herrera:1997plx,Mak:2001eb,Cardoso:2019rvt,Raposo:2018rjn,Becerra:2024wku} and references therein) or binary systems. However, this may not be so easy not only because of observational difficulties, but also because of an inherent limitation of our approach. Our mass formula \eqref{massshift} holds for static (eternal) local objects, i.e., objects with no local time-dependent dynamics. The only allowed time-dependence is the large-scale, cosmological one. This is a rather strong limitation. It excludes, for instance, binary systems in which the period of rotation changes over time due gravitational-wave emission. Incidentally, the inclusion of local time-dependent dynamics, together with the cosmological one in the system \eqref{systemalphadotzero}, seems to be a formidable task, unless the time-scale of the local dynamics is several orders of magnitude less than the cosmological one.

From the observational point of view, we expect the detection of the cosmological coupling of galaxies or other objects in the weak-field regime to be challenging. This is due to the uncertainties in measuring galactic masses, which are well above 10\%. In addition, it is quite difficult to find homogeneous samples at different redshifts. Therefore, the search for other physical parameters, which are sensitive to the cosmological mass shift, is of paramount importance. They might provide us with indirect proof of the cosmological coupling of objects in the weak-field regime.

\section*{Acknowledgements}
We thank Luca Amendola, Paolo Serra and Andrea Tarchi for fruitful discussions. DCR thanks Heidelberg University, Germany, and \textit{Universidade Federal do Rio de Janeiro}, Brazil, for hospitality and support. He also acknowledges partial support from \textit{Conselho Nacional de Desenvolvimento Científico e Tecnológico} (CNPq-Brazil) and \textit{Fundação de Amparo à Pesquisa e Inovação do Espírito Santo} (FAPES-Brazil).

\bigskip

\bibliography{Galaxies-coupling}

\begin{thebibliography}{79}%
\makeatletter
\providecommand \@ifxundefined [1]{%
 \@ifx{#1\undefined}
}%
\providecommand \@ifnum [1]{%
 \ifnum #1\expandafter \@firstoftwo
 \else \expandafter \@secondoftwo
 \fi
}%
\providecommand \@ifx [1]{%
 \ifx #1\expandafter \@firstoftwo
 \else \expandafter \@secondoftwo
 \fi
}%
\providecommand \natexlab [1]{#1}%
\providecommand \enquote  [1]{``#1''}%
\providecommand \bibnamefont  [1]{#1}%
\providecommand \bibfnamefont [1]{#1}%
\providecommand \citenamefont [1]{#1}%
\providecommand \href@noop [0]{\@secondoftwo}%
\providecommand \href [0]{\begingroup \@sanitize@url \@href}%
\providecommand \@href[1]{\@@startlink{#1}\@@href}%
\providecommand \@@href[1]{\endgroup#1\@@endlink}%
\providecommand \@sanitize@url [0]{\catcode `\\12\catcode `\$12\catcode `\&12\catcode `\#12\catcode `\^12\catcode `\_12\catcode `\%12\relax}%
\providecommand \@@startlink[1]{}%
\providecommand \@@endlink[0]{}%
\providecommand \url  [0]{\begingroup\@sanitize@url \@url }%
\providecommand \@url [1]{\endgroup\@href {#1}{\urlprefix }}%
\providecommand \urlprefix  [0]{URL }%
\providecommand \Eprint [0]{\href }%
\providecommand \doibase [0]{http://dx.doi.org/}%
\providecommand \selectlanguage [0]{\@gobble}%
\providecommand \bibinfo  [0]{\@secondoftwo}%
\providecommand \bibfield  [0]{\@secondoftwo}%
\providecommand \translation [1]{[#1]}%
\providecommand \BibitemOpen [0]{}%
\providecommand \bibitemStop [0]{}%
\providecommand \bibitemNoStop [0]{.\EOS\space}%
\providecommand \EOS [0]{\spacefactor3000\relax}%
\providecommand \BibitemShut  [1]{\csname bibitem#1\endcsname}%
\let\auto@bib@innerbib\@empty
\bibitem [{\citenamefont {Gao}\ and\ \citenamefont {Li}(2023)}]{Gao:2023keg}%
  \BibitemOpen
  \bibfield  {author} {\bibinfo {author} {\bibfnamefont {S.-J.}\ \bibnamefont {Gao}}\ and\ \bibinfo {author} {\bibfnamefont {X.-D.}\ \bibnamefont {Li}},\ }\href {\doibase 10.3847/1538-4357/ace890} {\bibfield  {journal} {\bibinfo  {journal} {Astrophys. J.}\ }\textbf {\bibinfo {volume} {956}},\ \bibinfo {pages} {128} (\bibinfo {year} {2023})},\ \Eprint {http://arxiv.org/abs/2307.10708} {arXiv:2307.10708 [astro-ph.HE]} \BibitemShut {NoStop}%
\bibitem [{\citenamefont {McVittie}(1933)}]{McVittie:1933zz}%
  \BibitemOpen
  \bibfield  {author} {\bibinfo {author} {\bibfnamefont {G.~C.}\ \bibnamefont {McVittie}},\ }\href {\doibase 10.1093/mnras/93.5.325} {\bibfield  {journal} {\bibinfo  {journal} {Mon. Not. Roy. Astron. Soc.}\ }\textbf {\bibinfo {volume} {93}},\ \bibinfo {pages} {325} (\bibinfo {year} {1933})}\BibitemShut {NoStop}%
\bibitem [{\citenamefont {Einstein}\ and\ \citenamefont {Straus}({\natexlab{a}})}]{Einstein:1945id}%
  \BibitemOpen
  \bibfield  {author} {\bibinfo {author} {\bibfnamefont {A.}~\bibnamefont {Einstein}}\ and\ \bibinfo {author} {\bibfnamefont {E.~G.}\ \bibnamefont {Straus}},\ }\href {\doibase 10.1103/RevModPhys.17.120} {\bibfield  {journal} {\bibinfo  {journal} {\href{http://dx.doi.org/10.1103/RevModPhys.17.120} {Rev. Mod. Phys. \textbf{17} (1945), 120-124}}\ } ({\natexlab{a}}),\ 10.1103/RevModPhys.17.120}\BibitemShut {NoStop}%
\bibitem [{\citenamefont {Einstein}\ and\ \citenamefont {Straus}({\natexlab{b}})}]{Einstein:1946zz}%
  \BibitemOpen
  \bibfield  {author} {\bibinfo {author} {\bibfnamefont {A.}~\bibnamefont {Einstein}}\ and\ \bibinfo {author} {\bibfnamefont {E.~G.}\ \bibnamefont {Straus}},\ }\href {\doibase 10.1103/RevModPhys.18.148} {\bibfield  {journal} {\bibinfo  {journal} {\href{http://dx.doi.org/10.1103/RevModPhys.18.148} {Rev. Mod. Phys. \textbf{18} (1946), 148-149}}\ } ({\natexlab{b}}),\ 10.1103/RevModPhys.18.148}\BibitemShut {NoStop}%
\bibitem [{\citenamefont {Pachner}(1963)}]{Pachner:1963zz}%
  \BibitemOpen
  \bibfield  {author} {\bibinfo {author} {\bibfnamefont {J.}~\bibnamefont {Pachner}},\ }\href {\doibase 10.1103/PhysRev.132.1837} {\bibfield  {journal} {\bibinfo  {journal} {Phys. Rev.}\ }\textbf {\bibinfo {volume} {132}},\ \bibinfo {pages} {1837} (\bibinfo {year} {1963})}\BibitemShut {NoStop}%
\bibitem [{\citenamefont {Dicke}\ and\ \citenamefont {Peebles}(1964)}]{dicke1964evolution}%
  \BibitemOpen
  \bibfield  {author} {\bibinfo {author} {\bibfnamefont {R.~H.}\ \bibnamefont {Dicke}}\ and\ \bibinfo {author} {\bibfnamefont {P.~J.~E.}\ \bibnamefont {Peebles}},\ }\href@noop {} {\bibfield  {journal} {\bibinfo  {journal} {Physical Review Letters}\ }\textbf {\bibinfo {volume} {12}},\ \bibinfo {pages} {435} (\bibinfo {year} {1964})}\BibitemShut {NoStop}%
\bibitem [{\citenamefont {Vaidya}\ and\ \citenamefont {Yashodhara}(1968)}]{Vaidya:1968zza}%
  \BibitemOpen
  \bibfield  {author} {\bibinfo {author} {\bibfnamefont {P.~C.}\ \bibnamefont {Vaidya}}\ and\ \bibinfo {author} {\bibfnamefont {P.~S.}\ \bibnamefont {Yashodhara}},\ }\href@noop {} {\bibfield  {journal} {\bibinfo  {journal} {Tensor (Japan)}\ }\textbf {\bibinfo {volume} {19}},\ \bibinfo {pages} {191} (\bibinfo {year} {1968})}\BibitemShut {NoStop}%
\bibitem [{\citenamefont {D'Eath}(1975)}]{DEath:1975jps}%
  \BibitemOpen
  \bibfield  {author} {\bibinfo {author} {\bibfnamefont {P.~D.}\ \bibnamefont {D'Eath}},\ }\href {\doibase 10.1103/PhysRevD.11.1387} {\bibfield  {journal} {\bibinfo  {journal} {Phys. Rev. D}\ }\textbf {\bibinfo {volume} {11}},\ \bibinfo {pages} {1387} (\bibinfo {year} {1975})}\BibitemShut {NoStop}%
\bibitem [{\citenamefont {Gautreau}(1984)}]{Gautreau:1984pny}%
  \BibitemOpen
  \bibfield  {author} {\bibinfo {author} {\bibfnamefont {R.}~\bibnamefont {Gautreau}},\ }\href {\doibase 10.1103/PhysRevD.29.198} {\bibfield  {journal} {\bibinfo  {journal} {Phys. Rev. D}\ }\textbf {\bibinfo {volume} {29}},\ \bibinfo {pages} {198} (\bibinfo {year} {1984})}\BibitemShut {NoStop}%
\bibitem [{\citenamefont {Cooperstock}\ \emph {et~al.}(1998)\citenamefont {Cooperstock}, \citenamefont {Faraoni},\ and\ \citenamefont {Vollick}}]{Cooperstock:1998ny}%
  \BibitemOpen
  \bibfield  {author} {\bibinfo {author} {\bibfnamefont {F.~I.}\ \bibnamefont {Cooperstock}}, \bibinfo {author} {\bibfnamefont {V.}~\bibnamefont {Faraoni}}, \ and\ \bibinfo {author} {\bibfnamefont {D.~N.}\ \bibnamefont {Vollick}},\ }\href {\doibase 10.1086/305956} {\bibfield  {journal} {\bibinfo  {journal} {Astrophys. J.}\ }\textbf {\bibinfo {volume} {503}},\ \bibinfo {pages} {61} (\bibinfo {year} {1998})},\ \Eprint {http://arxiv.org/abs/astro-ph/9803097} {arXiv:astro-ph/9803097} \BibitemShut {NoStop}%
\bibitem [{\citenamefont {Nayak}\ \emph {et~al.}(2001)\citenamefont {Nayak}, \citenamefont {MacCallum},\ and\ \citenamefont {Vishveshwara}}]{Nayak:2000mr}%
  \BibitemOpen
  \bibfield  {author} {\bibinfo {author} {\bibfnamefont {K.~R.}\ \bibnamefont {Nayak}}, \bibinfo {author} {\bibfnamefont {M.~A.~H.}\ \bibnamefont {MacCallum}}, \ and\ \bibinfo {author} {\bibfnamefont {C.~V.}\ \bibnamefont {Vishveshwara}},\ }\href {\doibase 10.1103/PhysRevD.63.024020} {\bibfield  {journal} {\bibinfo  {journal} {Phys. Rev. D}\ }\textbf {\bibinfo {volume} {63}},\ \bibinfo {pages} {024020} (\bibinfo {year} {2001})},\ \Eprint {http://arxiv.org/abs/gr-qc/0006040} {arXiv:gr-qc/0006040} \BibitemShut {NoStop}%
\bibitem [{\citenamefont {Baker}(2000)}]{Baker:2000yh}%
  \BibitemOpen
  \bibfield  {author} {\bibinfo {author} {\bibfnamefont {G.~A.}\ \bibnamefont {Baker}},\ }\href@noop {} {\  (\bibinfo {year} {2000})},\ \Eprint {http://arxiv.org/abs/astro-ph/0003152} {arXiv:astro-ph/0003152} \BibitemShut {NoStop}%
\bibitem [{\citenamefont {Bolen}\ \emph {et~al.}(2001)\citenamefont {Bolen}, \citenamefont {Bombelli},\ and\ \citenamefont {Puzio}}]{Bolen:2000dz}%
  \BibitemOpen
  \bibfield  {author} {\bibinfo {author} {\bibfnamefont {B.}~\bibnamefont {Bolen}}, \bibinfo {author} {\bibfnamefont {L.}~\bibnamefont {Bombelli}}, \ and\ \bibinfo {author} {\bibfnamefont {R.}~\bibnamefont {Puzio}},\ }\href {\doibase 10.1088/0264-9381/18/7/302} {\bibfield  {journal} {\bibinfo  {journal} {Class. Quant. Grav.}\ }\textbf {\bibinfo {volume} {18}},\ \bibinfo {pages} {1173} (\bibinfo {year} {2001})},\ \Eprint {http://arxiv.org/abs/gr-qc/0009018} {arXiv:gr-qc/0009018} \BibitemShut {NoStop}%
\bibitem [{\citenamefont {Dominguez}\ and\ \citenamefont {Gaite}(2001)}]{Dominguez:2001it}%
  \BibitemOpen
  \bibfield  {author} {\bibinfo {author} {\bibfnamefont {A.}~\bibnamefont {Dominguez}}\ and\ \bibinfo {author} {\bibfnamefont {J.~C.}\ \bibnamefont {Gaite}},\ }\href {\doibase 10.1209/epl/i2001-00437-y} {\bibfield  {journal} {\bibinfo  {journal} {EPL}\ }\textbf {\bibinfo {volume} {55}},\ \bibinfo {pages} {458} (\bibinfo {year} {2001})},\ \Eprint {http://arxiv.org/abs/astro-ph/0106199} {arXiv:astro-ph/0106199} \BibitemShut {NoStop}%
\bibitem [{\citenamefont {Ellis}(2002)}]{Ellis:2001cq}%
  \BibitemOpen
  \bibfield  {author} {\bibinfo {author} {\bibfnamefont {G.~F.~R.}\ \bibnamefont {Ellis}},\ }\href {\doibase 10.1142/S0217751X02011588} {\bibfield  {journal} {\bibinfo  {journal} {Int. J. Mod. Phys. A}\ }\textbf {\bibinfo {volume} {17}},\ \bibinfo {pages} {2667} (\bibinfo {year} {2002})},\ \Eprint {http://arxiv.org/abs/gr-qc/0102017} {arXiv:gr-qc/0102017} \BibitemShut {NoStop}%
\bibitem [{\citenamefont {Gao}\ and\ \citenamefont {Zhang}(2004)}]{Gao:2004cr}%
  \BibitemOpen
  \bibfield  {author} {\bibinfo {author} {\bibfnamefont {C.~J.}\ \bibnamefont {Gao}}\ and\ \bibinfo {author} {\bibfnamefont {S.~N.}\ \bibnamefont {Zhang}},\ }\href {\doibase 10.1016/j.physletb.2004.05.076} {\bibfield  {journal} {\bibinfo  {journal} {Phys. Lett. B}\ }\textbf {\bibinfo {volume} {595}},\ \bibinfo {pages} {28} (\bibinfo {year} {2004})},\ \Eprint {http://arxiv.org/abs/gr-qc/0407045} {arXiv:gr-qc/0407045} \BibitemShut {NoStop}%
\bibitem [{\citenamefont {Sheehan}\ and\ \citenamefont {Kriss}(2004)}]{Sheehan:2004wa}%
  \BibitemOpen
  \bibfield  {author} {\bibinfo {author} {\bibfnamefont {D.~P.}\ \bibnamefont {Sheehan}}\ and\ \bibinfo {author} {\bibfnamefont {V.~G.}\ \bibnamefont {Kriss}},\ }\href@noop {} {\  (\bibinfo {year} {2004})},\ \Eprint {http://arxiv.org/abs/astro-ph/0411299} {arXiv:astro-ph/0411299} \BibitemShut {NoStop}%
\bibitem [{\citenamefont {Nesseris}\ and\ \citenamefont {Perivolaropoulos}(2004)}]{Nesseris:2004uj}%
  \BibitemOpen
  \bibfield  {author} {\bibinfo {author} {\bibfnamefont {S.}~\bibnamefont {Nesseris}}\ and\ \bibinfo {author} {\bibfnamefont {L.}~\bibnamefont {Perivolaropoulos}},\ }\href {\doibase 10.1103/PhysRevD.70.123529} {\bibfield  {journal} {\bibinfo  {journal} {Phys. Rev. D}\ }\textbf {\bibinfo {volume} {70}},\ \bibinfo {pages} {123529} (\bibinfo {year} {2004})},\ \Eprint {http://arxiv.org/abs/astro-ph/0410309} {arXiv:astro-ph/0410309} \BibitemShut {NoStop}%
\bibitem [{\citenamefont {Sultana}\ and\ \citenamefont {Dyer}(2005)}]{Sultana:2005tp}%
  \BibitemOpen
  \bibfield  {author} {\bibinfo {author} {\bibfnamefont {J.}~\bibnamefont {Sultana}}\ and\ \bibinfo {author} {\bibfnamefont {C.~C.}\ \bibnamefont {Dyer}},\ }\href {\doibase 10.1007/s10714-005-0119-7} {\bibfield  {journal} {\bibinfo  {journal} {Gen. Rel. Grav.}\ }\textbf {\bibinfo {volume} {37}},\ \bibinfo {pages} {1347} (\bibinfo {year} {2005})}\BibitemShut {NoStop}%
\bibitem [{\citenamefont {Li}\ and\ \citenamefont {Wang}(2007)}]{Li:2006zh}%
  \BibitemOpen
  \bibfield  {author} {\bibinfo {author} {\bibfnamefont {Z.-H.}\ \bibnamefont {Li}}\ and\ \bibinfo {author} {\bibfnamefont {A.}~\bibnamefont {Wang}},\ }\href {\doibase 10.1142/S0217732307024048} {\bibfield  {journal} {\bibinfo  {journal} {Mod. Phys. Lett. A}\ }\textbf {\bibinfo {volume} {22}},\ \bibinfo {pages} {1663} (\bibinfo {year} {2007})},\ \Eprint {http://arxiv.org/abs/astro-ph/0607554} {arXiv:astro-ph/0607554} \BibitemShut {NoStop}%
\bibitem [{\citenamefont {Adkins}\ \emph {et~al.}(2007)\citenamefont {Adkins}, \citenamefont {McDonnell},\ and\ \citenamefont {Fell}}]{Adkins:2006kw}%
  \BibitemOpen
  \bibfield  {author} {\bibinfo {author} {\bibfnamefont {G.~S.}\ \bibnamefont {Adkins}}, \bibinfo {author} {\bibfnamefont {J.}~\bibnamefont {McDonnell}}, \ and\ \bibinfo {author} {\bibfnamefont {R.~N.}\ \bibnamefont {Fell}},\ }\href {\doibase 10.1103/PhysRevD.75.064011} {\bibfield  {journal} {\bibinfo  {journal} {Phys. Rev. D}\ }\textbf {\bibinfo {volume} {75}},\ \bibinfo {pages} {064011} (\bibinfo {year} {2007})},\ \Eprint {http://arxiv.org/abs/gr-qc/0612146} {arXiv:gr-qc/0612146} \BibitemShut {NoStop}%
\bibitem [{\citenamefont {McClure}\ and\ \citenamefont {Dyer}(2006)}]{McClure:2006kg}%
  \BibitemOpen
  \bibfield  {author} {\bibinfo {author} {\bibfnamefont {M.~L.}\ \bibnamefont {McClure}}\ and\ \bibinfo {author} {\bibfnamefont {C.~C.}\ \bibnamefont {Dyer}},\ }\href {\doibase 10.1088/0264-9381/23/6/008} {\bibfield  {journal} {\bibinfo  {journal} {Class. Quant. Grav.}\ }\textbf {\bibinfo {volume} {23}},\ \bibinfo {pages} {1971} (\bibinfo {year} {2006})}\BibitemShut {NoStop}%
\bibitem [{\citenamefont {Sereno}\ and\ \citenamefont {Jetzer}(2007)}]{Sereno:2007tt}%
  \BibitemOpen
  \bibfield  {author} {\bibinfo {author} {\bibfnamefont {M.}~\bibnamefont {Sereno}}\ and\ \bibinfo {author} {\bibfnamefont {P.}~\bibnamefont {Jetzer}},\ }\href {\doibase 10.1103/PhysRevD.75.064031} {\bibfield  {journal} {\bibinfo  {journal} {Phys. Rev. D}\ }\textbf {\bibinfo {volume} {75}},\ \bibinfo {pages} {064031} (\bibinfo {year} {2007})},\ \Eprint {http://arxiv.org/abs/astro-ph/0703121} {arXiv:astro-ph/0703121} \BibitemShut {NoStop}%
\bibitem [{\citenamefont {Faraoni}\ and\ \citenamefont {Jacques}(2007)}]{Faraoni:2007es}%
  \BibitemOpen
  \bibfield  {author} {\bibinfo {author} {\bibfnamefont {V.}~\bibnamefont {Faraoni}}\ and\ \bibinfo {author} {\bibfnamefont {A.}~\bibnamefont {Jacques}},\ }\href {\doibase 10.1103/PhysRevD.76.063510} {\bibfield  {journal} {\bibinfo  {journal} {Phys. Rev. D}\ }\textbf {\bibinfo {volume} {76}},\ \bibinfo {pages} {063510} (\bibinfo {year} {2007})},\ \Eprint {http://arxiv.org/abs/0707.1350} {arXiv:0707.1350 [gr-qc]} \BibitemShut {NoStop}%
\bibitem [{\citenamefont {Balaguera-Antolinez}\ and\ \citenamefont {Nowakowski}(2007)}]{Balaguera-Antolinez:2007csw}%
  \BibitemOpen
  \bibfield  {author} {\bibinfo {author} {\bibfnamefont {A.}~\bibnamefont {Balaguera-Antolinez}}\ and\ \bibinfo {author} {\bibfnamefont {M.}~\bibnamefont {Nowakowski}},\ }\href {\doibase 10.1088/0264-9381/24/10/013} {\bibfield  {journal} {\bibinfo  {journal} {Class. Quant. Grav.}\ }\textbf {\bibinfo {volume} {24}},\ \bibinfo {pages} {2677} (\bibinfo {year} {2007})},\ \Eprint {http://arxiv.org/abs/0704.1871} {arXiv:0704.1871 [gr-qc]} \BibitemShut {NoStop}%
\bibitem [{\citenamefont {Mashhoon}\ \emph {et~al.}(2007)\citenamefont {Mashhoon}, \citenamefont {Mobed},\ and\ \citenamefont {Singh}}]{Mashhoon:2007qm}%
  \BibitemOpen
  \bibfield  {author} {\bibinfo {author} {\bibfnamefont {B.}~\bibnamefont {Mashhoon}}, \bibinfo {author} {\bibfnamefont {N.}~\bibnamefont {Mobed}}, \ and\ \bibinfo {author} {\bibfnamefont {D.}~\bibnamefont {Singh}},\ }\href {\doibase 10.1088/0264-9381/24/20/008} {\bibfield  {journal} {\bibinfo  {journal} {Class. Quant. Grav.}\ }\textbf {\bibinfo {volume} {24}},\ \bibinfo {pages} {5031} (\bibinfo {year} {2007})},\ \Eprint {http://arxiv.org/abs/0705.1312} {arXiv:0705.1312 [gr-qc]} \BibitemShut {NoStop}%
\bibitem [{\citenamefont {Carrera}\ and\ \citenamefont {Giulini}(2010)}]{Carrera:2008pi}%
  \BibitemOpen
  \bibfield  {author} {\bibinfo {author} {\bibfnamefont {M.}~\bibnamefont {Carrera}}\ and\ \bibinfo {author} {\bibfnamefont {D.}~\bibnamefont {Giulini}},\ }\href {\doibase 10.1103/RevModPhys.82.169} {\bibfield  {journal} {\bibinfo  {journal} {Rev. Mod. Phys.}\ }\textbf {\bibinfo {volume} {82}},\ \bibinfo {pages} {169} (\bibinfo {year} {2010})},\ \Eprint {http://arxiv.org/abs/0810.2712} {arXiv:0810.2712 [gr-qc]} \BibitemShut {NoStop}%
\bibitem [{\citenamefont {Gao}\ \emph {et~al.}(2011)\citenamefont {Gao}, \citenamefont {Chen}, \citenamefont {Shen},\ and\ \citenamefont {Faraoni}}]{Gao:2011tq}%
  \BibitemOpen
  \bibfield  {author} {\bibinfo {author} {\bibfnamefont {C.}~\bibnamefont {Gao}}, \bibinfo {author} {\bibfnamefont {X.}~\bibnamefont {Chen}}, \bibinfo {author} {\bibfnamefont {Y.-G.}\ \bibnamefont {Shen}}, \ and\ \bibinfo {author} {\bibfnamefont {V.}~\bibnamefont {Faraoni}},\ }\href {\doibase 10.1103/PhysRevD.84.104047} {\bibfield  {journal} {\bibinfo  {journal} {Phys. Rev. D}\ }\textbf {\bibinfo {volume} {84}},\ \bibinfo {pages} {104047} (\bibinfo {year} {2011})},\ \Eprint {http://arxiv.org/abs/1110.6708} {arXiv:1110.6708 [gr-qc]} \BibitemShut {NoStop}%
\bibitem [{\citenamefont {Faraoni}\ \emph {et~al.}(2014)\citenamefont {Faraoni}, \citenamefont {Zambrano~Moreno},\ and\ \citenamefont {Prain}}]{Faraoni:2014nba}%
  \BibitemOpen
  \bibfield  {author} {\bibinfo {author} {\bibfnamefont {V.}~\bibnamefont {Faraoni}}, \bibinfo {author} {\bibfnamefont {A.~F.}\ \bibnamefont {Zambrano~Moreno}}, \ and\ \bibinfo {author} {\bibfnamefont {A.}~\bibnamefont {Prain}},\ }\href {\doibase 10.1103/PhysRevD.89.103514} {\bibfield  {journal} {\bibinfo  {journal} {Phys. Rev. D}\ }\textbf {\bibinfo {volume} {89}},\ \bibinfo {pages} {103514} (\bibinfo {year} {2014})},\ \Eprint {http://arxiv.org/abs/1404.3929} {arXiv:1404.3929 [gr-qc]} \BibitemShut {NoStop}%
\bibitem [{\citenamefont {Kopeikin}(2015)}]{Kopeikin:2014qna}%
  \BibitemOpen
  \bibfield  {author} {\bibinfo {author} {\bibfnamefont {S.}~\bibnamefont {Kopeikin}},\ }\href {\doibase 10.1140/epjp/i2015-15011-y} {\bibfield  {journal} {\bibinfo  {journal} {Eur. Phys. J. Plus}\ }\textbf {\bibinfo {volume} {130}},\ \bibinfo {pages} {11} (\bibinfo {year} {2015})},\ \Eprint {http://arxiv.org/abs/1407.6667} {arXiv:1407.6667 [astro-ph.CO]} \BibitemShut {NoStop}%
\bibitem [{\citenamefont {Faraoni}\ \emph {et~al.}(2015)\citenamefont {Faraoni}, \citenamefont {Lapierre-L\'eonard},\ and\ \citenamefont {Prain}}]{Faraoni:2015saa}%
  \BibitemOpen
  \bibfield  {author} {\bibinfo {author} {\bibfnamefont {V.}~\bibnamefont {Faraoni}}, \bibinfo {author} {\bibfnamefont {M.}~\bibnamefont {Lapierre-L\'eonard}}, \ and\ \bibinfo {author} {\bibfnamefont {A.}~\bibnamefont {Prain}},\ }\href {\doibase 10.1088/1475-7516/2015/10/013} {\bibfield  {journal} {\bibinfo  {journal} {JCAP}\ }\textbf {\bibinfo {volume} {10}},\ \bibinfo {pages} {013} (\bibinfo {year} {2015})},\ \Eprint {http://arxiv.org/abs/1508.01725} {arXiv:1508.01725 [gr-qc]} \BibitemShut {NoStop}%
\bibitem [{\citenamefont {Mello}\ \emph {et~al.}(2017)\citenamefont {Mello}, \citenamefont {Maciel},\ and\ \citenamefont {Zanchin}}]{Mello:2016irl}%
  \BibitemOpen
  \bibfield  {author} {\bibinfo {author} {\bibfnamefont {M.~M.~C.}\ \bibnamefont {Mello}}, \bibinfo {author} {\bibfnamefont {A.}~\bibnamefont {Maciel}}, \ and\ \bibinfo {author} {\bibfnamefont {V.~T.}\ \bibnamefont {Zanchin}},\ }\href {\doibase 10.1103/PhysRevD.95.084031} {\bibfield  {journal} {\bibinfo  {journal} {Phys. Rev. D}\ }\textbf {\bibinfo {volume} {95}},\ \bibinfo {pages} {084031} (\bibinfo {year} {2017})},\ \Eprint {http://arxiv.org/abs/1611.05077} {arXiv:1611.05077 [gr-qc]} \BibitemShut {NoStop}%
\bibitem [{\citenamefont {Faraoni}(2018)}]{Faraoni:2018xwo}%
  \BibitemOpen
  \bibfield  {author} {\bibinfo {author} {\bibfnamefont {V.}~\bibnamefont {Faraoni}},\ }\href {\doibase 10.3390/universe4100109} {\bibfield  {journal} {\bibinfo  {journal} {Universe}\ }\textbf {\bibinfo {volume} {4}},\ \bibinfo {pages} {109} (\bibinfo {year} {2018})},\ \Eprint {http://arxiv.org/abs/1810.04667} {arXiv:1810.04667 [gr-qc]} \BibitemShut {NoStop}%
\bibitem [{\citenamefont {Guariento}\ \emph {et~al.}(2019)\citenamefont {Guariento}, \citenamefont {Maciel}, \citenamefont {Mello},\ and\ \citenamefont {Zanchin}}]{Guariento:2019ock}%
  \BibitemOpen
  \bibfield  {author} {\bibinfo {author} {\bibfnamefont {D.~C.}\ \bibnamefont {Guariento}}, \bibinfo {author} {\bibfnamefont {A.}~\bibnamefont {Maciel}}, \bibinfo {author} {\bibfnamefont {M.~M.~C.}\ \bibnamefont {Mello}}, \ and\ \bibinfo {author} {\bibfnamefont {V.~T.}\ \bibnamefont {Zanchin}},\ }\href {\doibase 10.1103/PhysRevD.100.104050} {\bibfield  {journal} {\bibinfo  {journal} {Phys. Rev. D}\ }\textbf {\bibinfo {volume} {100}},\ \bibinfo {pages} {104050} (\bibinfo {year} {2019})},\ \Eprint {http://arxiv.org/abs/1908.04961} {arXiv:1908.04961 [gr-qc]} \BibitemShut {NoStop}%
\bibitem [{\citenamefont {Spengler}\ \emph {et~al.}(2022)\citenamefont {Spengler}, \citenamefont {Belenchia}, \citenamefont {R\"atzel},\ and\ \citenamefont {Braun}}]{Spengler:2021vxy}%
  \BibitemOpen
  \bibfield  {author} {\bibinfo {author} {\bibfnamefont {F.}~\bibnamefont {Spengler}}, \bibinfo {author} {\bibfnamefont {A.}~\bibnamefont {Belenchia}}, \bibinfo {author} {\bibfnamefont {D.}~\bibnamefont {R\"atzel}}, \ and\ \bibinfo {author} {\bibfnamefont {D.}~\bibnamefont {Braun}},\ }\href {\doibase 10.1088/1361-6382/ac4954} {\bibfield  {journal} {\bibinfo  {journal} {Class. Quant. Grav.}\ }\textbf {\bibinfo {volume} {39}},\ \bibinfo {pages} {055005} (\bibinfo {year} {2022})},\ \Eprint {http://arxiv.org/abs/2109.03280} {arXiv:2109.03280 [gr-qc]} \BibitemShut {NoStop}%
\bibitem [{\citenamefont {Agatsuma}(2022)}]{Agatsuma:2022ewd}%
  \BibitemOpen
  \bibfield  {author} {\bibinfo {author} {\bibfnamefont {K.}~\bibnamefont {Agatsuma}},\ }\href {\doibase 10.1016/j.dark.2022.101134} {\bibfield  {journal} {\bibinfo  {journal} {Phys. Dark Univ.}\ }\textbf {\bibinfo {volume} {38}},\ \bibinfo {pages} {101134} (\bibinfo {year} {2022})},\ \Eprint {http://arxiv.org/abs/2211.15668} {arXiv:2211.15668 [gr-qc]} \BibitemShut {NoStop}%
\bibitem [{\citenamefont {Croker}\ and\ \citenamefont {Weiner}(2019)}]{Croker:2019mup}%
  \BibitemOpen
  \bibfield  {author} {\bibinfo {author} {\bibfnamefont {K.~S.}\ \bibnamefont {Croker}}\ and\ \bibinfo {author} {\bibfnamefont {J.~L.}\ \bibnamefont {Weiner}},\ }\href {\doibase 10.3847/1538-4357/ab32da} {\bibfield  {journal} {\bibinfo  {journal} {Astrophys. J.}\ }\textbf {\bibinfo {volume} {882}},\ \bibinfo {pages} {19} (\bibinfo {year} {2019})},\ \Eprint {http://arxiv.org/abs/2107.06643} {arXiv:2107.06643 [gr-qc]} \BibitemShut {NoStop}%
\bibitem [{\citenamefont {Croker}\ \emph {et~al.}(2020{\natexlab{a}})\citenamefont {Croker}, \citenamefont {Nishimura},\ and\ \citenamefont {Farrah}}]{Croker:2020}%
  \BibitemOpen
  \bibfield  {author} {\bibinfo {author} {\bibfnamefont {K.~S.}\ \bibnamefont {Croker}}, \bibinfo {author} {\bibfnamefont {K.~A.}\ \bibnamefont {Nishimura}}, \ and\ \bibinfo {author} {\bibfnamefont {D.}~\bibnamefont {Farrah}},\ }\href {\doibase 10.3847/1538-4357/ab5aff} {\bibfield  {journal} {\bibinfo  {journal} {\href{http://dx.doi.org/10.3847/1538-4357/ab5aff} {Astrophys. J. \textbf{889} (2020) no.2, 115}}\ }\textbf {\bibinfo {volume} {889}},\ \bibinfo {pages} {115} (\bibinfo {year} {2020}{\natexlab{a}})}\BibitemShut {NoStop}%
\bibitem [{\citenamefont {Croker}\ \emph {et~al.}(2020{\natexlab{b}})\citenamefont {Croker}, \citenamefont {Runburg},\ and\ \citenamefont {Farrah}}]{Croker:2020plg}%
  \BibitemOpen
  \bibfield  {author} {\bibinfo {author} {\bibfnamefont {K.~S.}\ \bibnamefont {Croker}}, \bibinfo {author} {\bibfnamefont {J.}~\bibnamefont {Runburg}}, \ and\ \bibinfo {author} {\bibfnamefont {D.}~\bibnamefont {Farrah}},\ }\href {\doibase 10.3847/1538-4357/abad2f} {\bibfield  {journal} {\bibinfo  {journal} {Astrophys. J.}\ }\textbf {\bibinfo {volume} {900}},\ \bibinfo {pages} {57} (\bibinfo {year} {2020}{\natexlab{b}})}\BibitemShut {NoStop}%
\bibitem [{\citenamefont {Farrah}\ \emph {et~al.}(2023)\citenamefont {Farrah}, \citenamefont {Croker}, \citenamefont {Zevin}, \citenamefont {Tarlé}, \citenamefont {Faraoni}, \citenamefont {Petty}, \citenamefont {Afonso}, \citenamefont {Fernandez}, \citenamefont {Nishimura}, \citenamefont {Pearson}, \citenamefont {Wang}, \citenamefont {Clements}, \citenamefont {Efstathiou}, \citenamefont {Hatziminaoglou}, \citenamefont {Lacy}, \citenamefont {McPartland}, \citenamefont {Pitchford}, \citenamefont {Sakai},\ and\ \citenamefont {Weiner}}]{Farrah:2023opk}%
  \BibitemOpen
  \bibfield  {author} {\bibinfo {author} {\bibfnamefont {D.}~\bibnamefont {Farrah}}, \bibinfo {author} {\bibfnamefont {K.~S.}\ \bibnamefont {Croker}}, \bibinfo {author} {\bibfnamefont {M.}~\bibnamefont {Zevin}}, \bibinfo {author} {\bibfnamefont {G.}~\bibnamefont {Tarlé}}, \bibinfo {author} {\bibfnamefont {V.}~\bibnamefont {Faraoni}}, \bibinfo {author} {\bibfnamefont {S.}~\bibnamefont {Petty}}, \bibinfo {author} {\bibfnamefont {J.}~\bibnamefont {Afonso}}, \bibinfo {author} {\bibfnamefont {N.}~\bibnamefont {Fernandez}}, \bibinfo {author} {\bibfnamefont {K.~A.}\ \bibnamefont {Nishimura}}, \bibinfo {author} {\bibfnamefont {C.}~\bibnamefont {Pearson}}, \bibinfo {author} {\bibfnamefont {L.}~\bibnamefont {Wang}}, \bibinfo {author} {\bibfnamefont {D.~L.}\ \bibnamefont {Clements}}, \bibinfo {author} {\bibfnamefont {A.}~\bibnamefont {Efstathiou}}, \bibinfo {author} {\bibfnamefont {E.}~\bibnamefont {Hatziminaoglou}}, \bibinfo {author} {\bibfnamefont {M.}~\bibnamefont {Lacy}}, \bibinfo {author} {\bibfnamefont
  {C.}~\bibnamefont {McPartland}}, \bibinfo {author} {\bibfnamefont {L.~K.}\ \bibnamefont {Pitchford}}, \bibinfo {author} {\bibfnamefont {N.}~\bibnamefont {Sakai}}, \ and\ \bibinfo {author} {\bibfnamefont {J.}~\bibnamefont {Weiner}},\ }\href {\doibase 10.3847/2041-8213/acb704} {\bibfield  {journal} {\bibinfo  {journal} {The Astrophysical Journal Letters}\ }\textbf {\bibinfo {volume} {944}},\ \bibinfo {pages} {L31} (\bibinfo {year} {2023})}\BibitemShut {NoStop}%
\bibitem [{\citenamefont {Mistele}(2023)}]{Mistele:2023fds}%
  \BibitemOpen
  \bibfield  {author} {\bibinfo {author} {\bibfnamefont {T.}~\bibnamefont {Mistele}},\ }\href {\doibase 10.3847/2515-5172/acd767} {\bibfield  {journal} {\bibinfo  {journal} {Res. Notes AAS}\ }\textbf {\bibinfo {volume} {7}},\ \bibinfo {pages} {101} (\bibinfo {year} {2023})},\ \Eprint {http://arxiv.org/abs/2304.09817} {arXiv:2304.09817 [gr-qc]} \BibitemShut {NoStop}%
\bibitem [{\citenamefont {Wang}\ and\ \citenamefont {Wang}(2023)}]{Wang:2023aqe}%
  \BibitemOpen
  \bibfield  {author} {\bibinfo {author} {\bibfnamefont {Y.}~\bibnamefont {Wang}}\ and\ \bibinfo {author} {\bibfnamefont {Z.}~\bibnamefont {Wang}},\ }\href@noop {} {\  (\bibinfo {year} {2023})},\ \Eprint {http://arxiv.org/abs/2304.01059} {arXiv:2304.01059 [gr-qc]} \BibitemShut {NoStop}%
\bibitem [{\citenamefont {Gaur}\ and\ \citenamefont {Visser}(2023)}]{Gaur:2023hmk}%
  \BibitemOpen
  \bibfield  {author} {\bibinfo {author} {\bibfnamefont {R.}~\bibnamefont {Gaur}}\ and\ \bibinfo {author} {\bibfnamefont {M.}~\bibnamefont {Visser}},\ }\href@noop {} {\  (\bibinfo {year} {2023})},\ \Eprint {http://arxiv.org/abs/2308.07374} {arXiv:2308.07374 [gr-qc]} \BibitemShut {NoStop}%
\bibitem [{\citenamefont {Parnovsky}(2023)}]{Parnovsky:2023wkc}%
  \BibitemOpen
  \bibfield  {author} {\bibinfo {author} {\bibfnamefont {S.~L.}\ \bibnamefont {Parnovsky}},\ }\href@noop {} {\  (\bibinfo {year} {2023})},\ \Eprint {http://arxiv.org/abs/2302.13333} {arXiv:2302.13333 [gr-qc]} \BibitemShut {NoStop}%
\bibitem [{\citenamefont {Avelino}(2023)}]{Avelino:2023rac}%
  \BibitemOpen
  \bibfield  {author} {\bibinfo {author} {\bibfnamefont {P.~P.}\ \bibnamefont {Avelino}},\ }\href {\doibase 10.1088/1475-7516/2023/08/005} {\bibfield  {journal} {\bibinfo  {journal} {JCAP}\ }\textbf {\bibinfo {volume} {08}},\ \bibinfo {pages} {005} (\bibinfo {year} {2023})},\ \Eprint {http://arxiv.org/abs/2303.06630} {arXiv:2303.06630 [gr-qc]} \BibitemShut {NoStop}%
\bibitem [{\citenamefont {Dahal}\ \emph {et~al.}(2023)\citenamefont {Dahal}, \citenamefont {Maharana}, \citenamefont {Simovic}, \citenamefont {Soranidis},\ and\ \citenamefont {Terno}}]{Dahal:2023hzo}%
  \BibitemOpen
  \bibfield  {author} {\bibinfo {author} {\bibfnamefont {P.~K.}\ \bibnamefont {Dahal}}, \bibinfo {author} {\bibfnamefont {S.}~\bibnamefont {Maharana}}, \bibinfo {author} {\bibfnamefont {F.}~\bibnamefont {Simovic}}, \bibinfo {author} {\bibfnamefont {I.}~\bibnamefont {Soranidis}}, \ and\ \bibinfo {author} {\bibfnamefont {D.~R.}\ \bibnamefont {Terno}},\ }\href@noop {} {\  (\bibinfo {year} {2023})},\ \Eprint {http://arxiv.org/abs/2312.16804} {arXiv:2312.16804 [gr-qc]} \BibitemShut {NoStop}%
\bibitem [{\citenamefont {Cadoni}\ \emph {et~al.}(2023)\citenamefont {Cadoni}, \citenamefont {Sanna}, \citenamefont {Pitzalis}, \citenamefont {Banerjee}, \citenamefont {Murgia}, \citenamefont {Hazra},\ and\ \citenamefont {Branchesi}}]{Cadoni:2023lum}%
  \BibitemOpen
  \bibfield  {author} {\bibinfo {author} {\bibfnamefont {M.}~\bibnamefont {Cadoni}}, \bibinfo {author} {\bibfnamefont {A.~P.}\ \bibnamefont {Sanna}}, \bibinfo {author} {\bibfnamefont {M.}~\bibnamefont {Pitzalis}}, \bibinfo {author} {\bibfnamefont {B.}~\bibnamefont {Banerjee}}, \bibinfo {author} {\bibfnamefont {R.}~\bibnamefont {Murgia}}, \bibinfo {author} {\bibfnamefont {N.}~\bibnamefont {Hazra}}, \ and\ \bibinfo {author} {\bibfnamefont {M.}~\bibnamefont {Branchesi}},\ }\href {\doibase 10.1088/1475-7516/2023/11/007} {\bibfield  {journal} {\bibinfo  {journal} {JCAP}\ }\textbf {\bibinfo {volume} {11}},\ \bibinfo {pages} {007} (\bibinfo {year} {2023})},\ \Eprint {http://arxiv.org/abs/2306.11588} {arXiv:2306.11588 [gr-qc]} \BibitemShut {NoStop}%
\bibitem [{\citenamefont {Cadoni}\ \emph {et~al.}(2024)\citenamefont {Cadoni}, \citenamefont {Murgia}, \citenamefont {Pitzalis},\ and\ \citenamefont {Sanna}}]{Cadoni:2023lqe}%
  \BibitemOpen
  \bibfield  {author} {\bibinfo {author} {\bibfnamefont {M.}~\bibnamefont {Cadoni}}, \bibinfo {author} {\bibfnamefont {R.}~\bibnamefont {Murgia}}, \bibinfo {author} {\bibfnamefont {M.}~\bibnamefont {Pitzalis}}, \ and\ \bibinfo {author} {\bibfnamefont {A.~P.}\ \bibnamefont {Sanna}},\ }\href {\doibase 10.1088/1475-7516/2024/03/026} {\bibfield  {journal} {\bibinfo  {journal} {JCAP}\ }\textbf {\bibinfo {volume} {03}},\ \bibinfo {pages} {026} (\bibinfo {year} {2024})},\ \Eprint {http://arxiv.org/abs/2309.16444} {arXiv:2309.16444 [gr-qc]} \BibitemShut {NoStop}%
\bibitem [{\citenamefont {Rodriguez}(2023)}]{Rodriguez:2023gaa}%
  \BibitemOpen
  \bibfield  {author} {\bibinfo {author} {\bibfnamefont {C.~L.}\ \bibnamefont {Rodriguez}},\ }\href {\doibase 10.3847/2041-8213/acc9b6} {\bibfield  {journal} {\bibinfo  {journal} {Astrophys. J. Lett.}\ }\textbf {\bibinfo {volume} {947}},\ \bibinfo {pages} {L12} (\bibinfo {year} {2023})},\ \Eprint {http://arxiv.org/abs/2302.12386} {arXiv:2302.12386 [astro-ph.CO]} \BibitemShut {NoStop}%
\bibitem [{\citenamefont {Andrae}\ and\ \citenamefont {El-Badry}(2023)}]{Andrae:2023wge}%
  \BibitemOpen
  \bibfield  {author} {\bibinfo {author} {\bibfnamefont {R.}~\bibnamefont {Andrae}}\ and\ \bibinfo {author} {\bibfnamefont {K.}~\bibnamefont {El-Badry}},\ }\href {\doibase 10.1051/0004-6361/202346350} {\bibfield  {journal} {\bibinfo  {journal} {Astron. Astrophys.}\ }\textbf {\bibinfo {volume} {673}},\ \bibinfo {pages} {L10} (\bibinfo {year} {2023})},\ \Eprint {http://arxiv.org/abs/2305.01307} {arXiv:2305.01307 [astro-ph.CO]} \BibitemShut {NoStop}%
\bibitem [{\citenamefont {Lei}\ \emph {et~al.}(2024)\citenamefont {Lei}, \citenamefont {Zu}, \citenamefont {Yuan}, \citenamefont {Shen}, \citenamefont {Wang}, \citenamefont {Wang}, \citenamefont {Su}, \citenamefont {Ren}, \citenamefont {Tang},\ and\ \citenamefont {Zhou}}]{Lei:2023mke}%
  \BibitemOpen
  \bibfield  {author} {\bibinfo {author} {\bibfnamefont {L.}~\bibnamefont {Lei}}, \bibinfo {author} {\bibfnamefont {L.}~\bibnamefont {Zu}}, \bibinfo {author} {\bibfnamefont {G.-W.}\ \bibnamefont {Yuan}}, \bibinfo {author} {\bibfnamefont {Z.-Q.}\ \bibnamefont {Shen}}, \bibinfo {author} {\bibfnamefont {Y.-Y.}\ \bibnamefont {Wang}}, \bibinfo {author} {\bibfnamefont {Y.-Z.}\ \bibnamefont {Wang}}, \bibinfo {author} {\bibfnamefont {Z.-B.}\ \bibnamefont {Su}}, \bibinfo {author} {\bibfnamefont {W.-k.}\ \bibnamefont {Ren}}, \bibinfo {author} {\bibfnamefont {S.-P.}\ \bibnamefont {Tang}}, \ and\ \bibinfo {author} {\bibfnamefont {H.}~\bibnamefont {Zhou}},\ }\href {\doibase 10.1007/s11433-023-2233-2} {\bibfield  {journal} {\bibinfo  {journal} {Sci. China Phys. Mech. Astron.}\ }\textbf {\bibinfo {volume} {67}},\ \bibinfo {pages} {229811} (\bibinfo {year} {2024})},\ \Eprint {http://arxiv.org/abs/2305.03408} {arXiv:2305.03408 [astro-ph.CO]} \BibitemShut {NoStop}%
\bibitem [{\citenamefont {Amendola}\ \emph {et~al.}(2024)\citenamefont {Amendola}, \citenamefont {Rodrigues}, \citenamefont {Kumar},\ and\ \citenamefont {Quartin}}]{Amendola:2023ays}%
  \BibitemOpen
  \bibfield  {author} {\bibinfo {author} {\bibfnamefont {L.}~\bibnamefont {Amendola}}, \bibinfo {author} {\bibfnamefont {D.~C.}\ \bibnamefont {Rodrigues}}, \bibinfo {author} {\bibfnamefont {S.}~\bibnamefont {Kumar}}, \ and\ \bibinfo {author} {\bibfnamefont {M.}~\bibnamefont {Quartin}},\ }\href {\doibase 10.1093/mnras/stae143} {\bibfield  {journal} {\bibinfo  {journal} {Mon. Not. Roy. Astron. Soc.}\ }\textbf {\bibinfo {volume} {528}},\ \bibinfo {pages} {2377} (\bibinfo {year} {2024})},\ \Eprint {http://arxiv.org/abs/2307.02474} {arXiv:2307.02474 [astro-ph.CO]} \BibitemShut {NoStop}%
\bibitem [{\citenamefont {Lacy}\ \emph {et~al.}(2024)\citenamefont {Lacy}, \citenamefont {Engholm}, \citenamefont {Farrah},\ and\ \citenamefont {Ejercito}}]{Lacy:2023kbb}%
  \BibitemOpen
  \bibfield  {author} {\bibinfo {author} {\bibfnamefont {M.}~\bibnamefont {Lacy}}, \bibinfo {author} {\bibfnamefont {A.}~\bibnamefont {Engholm}}, \bibinfo {author} {\bibfnamefont {D.}~\bibnamefont {Farrah}}, \ and\ \bibinfo {author} {\bibfnamefont {K.}~\bibnamefont {Ejercito}},\ }\href {\doibase 10.3847/2041-8213/ad1b5f} {\bibfield  {journal} {\bibinfo  {journal} {Astrophys. J. Lett.}\ }\textbf {\bibinfo {volume} {961}},\ \bibinfo {pages} {L33} (\bibinfo {year} {2024})},\ \Eprint {http://arxiv.org/abs/2312.12344} {arXiv:2312.12344 [astro-ph.CO]} \BibitemShut {NoStop}%
\bibitem [{\citenamefont {Cadoni}\ and\ \citenamefont {Sanna}(2021{\natexlab{a}})}]{Cadoni:2020jxe}%
  \BibitemOpen
  \bibfield  {author} {\bibinfo {author} {\bibfnamefont {M.}~\bibnamefont {Cadoni}}\ and\ \bibinfo {author} {\bibfnamefont {A.~P.}\ \bibnamefont {Sanna}},\ }\href {\doibase 10.1142/S0217751X21501566} {\bibfield  {journal} {\bibinfo  {journal} {Int. J. Mod. Phys. A}\ }\textbf {\bibinfo {volume} {36}},\ \bibinfo {pages} {2150156} (\bibinfo {year} {2021}{\natexlab{a}})},\ \Eprint {http://arxiv.org/abs/2012.08335} {arXiv:2012.08335 [gr-qc]} \BibitemShut {NoStop}%
\bibitem [{\citenamefont {Croker}\ \emph {et~al.}(2021)\citenamefont {Croker}, \citenamefont {Zevin}, \citenamefont {Farrah}, \citenamefont {Nishimura},\ and\ \citenamefont {Tarle}}]{Croker:2021duf}%
  \BibitemOpen
  \bibfield  {author} {\bibinfo {author} {\bibfnamefont {K.~S.}\ \bibnamefont {Croker}}, \bibinfo {author} {\bibfnamefont {M.~J.}\ \bibnamefont {Zevin}}, \bibinfo {author} {\bibfnamefont {D.}~\bibnamefont {Farrah}}, \bibinfo {author} {\bibfnamefont {K.~A.}\ \bibnamefont {Nishimura}}, \ and\ \bibinfo {author} {\bibfnamefont {G.}~\bibnamefont {Tarle}},\ }\href {\doibase 10.3847/2041-8213/ac2fad} {\bibfield  {journal} {\bibinfo  {journal} {Astrophys. J. Lett.}\ }\textbf {\bibinfo {volume} {921}},\ \bibinfo {pages} {L22} (\bibinfo {year} {2021})},\ \Eprint {http://arxiv.org/abs/2109.08146} {arXiv:2109.08146 [gr-qc]} \BibitemShut {NoStop}%
\bibitem [{\citenamefont {Croker}\ \emph {et~al.}(2019)\citenamefont {Croker}, \citenamefont {Nishimura},\ and\ \citenamefont {Farrah}}]{Croker:2019kje}%
  \BibitemOpen
  \bibfield  {author} {\bibinfo {author} {\bibfnamefont {K.}~\bibnamefont {Croker}}, \bibinfo {author} {\bibfnamefont {K.}~\bibnamefont {Nishimura}}, \ and\ \bibinfo {author} {\bibfnamefont {D.}~\bibnamefont {Farrah}},\ }\href {\doibase 10.3847/1538-4357/ab5aff} {\  (\bibinfo {year} {2019}),\ 10.3847/1538-4357/ab5aff},\ \Eprint {http://arxiv.org/abs/1904.03781} {arXiv:1904.03781 [astro-ph.CO]} \BibitemShut {NoStop}%
\bibitem [{\citenamefont {Cadoni}\ \emph {et~al.}(2022)\citenamefont {Cadoni}, \citenamefont {Oi},\ and\ \citenamefont {Sanna}}]{Cadoni:2022chn}%
  \BibitemOpen
  \bibfield  {author} {\bibinfo {author} {\bibfnamefont {M.}~\bibnamefont {Cadoni}}, \bibinfo {author} {\bibfnamefont {M.}~\bibnamefont {Oi}}, \ and\ \bibinfo {author} {\bibfnamefont {A.~P.}\ \bibnamefont {Sanna}},\ }\href {\doibase 10.1103/PhysRevD.106.024030} {\bibfield  {journal} {\bibinfo  {journal} {Phys. Rev. D}\ }\textbf {\bibinfo {volume} {106}},\ \bibinfo {pages} {024030} (\bibinfo {year} {2022})},\ \Eprint {http://arxiv.org/abs/2204.09444} {arXiv:2204.09444 [gr-qc]} \BibitemShut {NoStop}%
\bibitem [{\citenamefont {Cadoni}\ \emph {et~al.}(2018)\citenamefont {Cadoni}, \citenamefont {Casadio}, \citenamefont {Giusti}, \citenamefont {M\"uck},\ and\ \citenamefont {Tuveri}}]{Cadoni:2017evg}%
  \BibitemOpen
  \bibfield  {author} {\bibinfo {author} {\bibfnamefont {M.}~\bibnamefont {Cadoni}}, \bibinfo {author} {\bibfnamefont {R.}~\bibnamefont {Casadio}}, \bibinfo {author} {\bibfnamefont {A.}~\bibnamefont {Giusti}}, \bibinfo {author} {\bibfnamefont {W.}~\bibnamefont {M\"uck}}, \ and\ \bibinfo {author} {\bibfnamefont {M.}~\bibnamefont {Tuveri}},\ }\href {\doibase 10.1016/j.physletb.2017.11.058} {\bibfield  {journal} {\bibinfo  {journal} {Phys. Lett. B}\ }\textbf {\bibinfo {volume} {776}},\ \bibinfo {pages} {242} (\bibinfo {year} {2018})},\ \Eprint {http://arxiv.org/abs/1707.09945} {arXiv:1707.09945 [gr-qc]} \BibitemShut {NoStop}%
\bibitem [{\citenamefont {Tuveri}\ and\ \citenamefont {Cadoni}(2019{\natexlab{a}})}]{Tuveri:2019uej}%
  \BibitemOpen
  \bibfield  {author} {\bibinfo {author} {\bibfnamefont {M.}~\bibnamefont {Tuveri}}\ and\ \bibinfo {author} {\bibfnamefont {M.}~\bibnamefont {Cadoni}},\ }\href@noop {} {\  (\bibinfo {year} {2019}{\natexlab{a}})},\ \Eprint {http://arxiv.org/abs/1904.08209} {arXiv:1904.08209 [gr-qc]} \BibitemShut {NoStop}%
\bibitem [{\citenamefont {Tuveri}\ and\ \citenamefont {Cadoni}(2019{\natexlab{b}})}]{Tuveri:2019zor}%
  \BibitemOpen
  \bibfield  {author} {\bibinfo {author} {\bibfnamefont {M.}~\bibnamefont {Tuveri}}\ and\ \bibinfo {author} {\bibfnamefont {M.}~\bibnamefont {Cadoni}},\ }\href {\doibase 10.1103/PhysRevD.100.024029} {\bibfield  {journal} {\bibinfo  {journal} {Phys. Rev. D}\ }\textbf {\bibinfo {volume} {100}},\ \bibinfo {pages} {024029} (\bibinfo {year} {2019}{\natexlab{b}})},\ \Eprint {http://arxiv.org/abs/1904.11835} {arXiv:1904.11835 [gr-qc]} \BibitemShut {NoStop}%
\bibitem [{\citenamefont {Cadoni}\ \emph {et~al.}(2020)\citenamefont {Cadoni}, \citenamefont {Sanna},\ and\ \citenamefont {Tuveri}}]{Cadoni:2020izk}%
  \BibitemOpen
  \bibfield  {author} {\bibinfo {author} {\bibfnamefont {M.}~\bibnamefont {Cadoni}}, \bibinfo {author} {\bibfnamefont {A.~P.}\ \bibnamefont {Sanna}}, \ and\ \bibinfo {author} {\bibfnamefont {M.}~\bibnamefont {Tuveri}},\ }\href {\doibase 10.1103/PhysRevD.102.023514} {\bibfield  {journal} {\bibinfo  {journal} {Phys. Rev. D}\ }\textbf {\bibinfo {volume} {102}},\ \bibinfo {pages} {023514} (\bibinfo {year} {2020})},\ \Eprint {http://arxiv.org/abs/2002.06988} {arXiv:2002.06988 [gr-qc]} \BibitemShut {NoStop}%
\bibitem [{\citenamefont {Cadoni}\ and\ \citenamefont {Sanna}(2021{\natexlab{b}})}]{Cadoni:2021zsl}%
  \BibitemOpen
  \bibfield  {author} {\bibinfo {author} {\bibfnamefont {M.}~\bibnamefont {Cadoni}}\ and\ \bibinfo {author} {\bibfnamefont {A.~P.}\ \bibnamefont {Sanna}},\ }\href {\doibase 10.1088/1361-6382/abfd92} {\bibfield  {journal} {\bibinfo  {journal} {Class. Quant. Grav.}\ }\textbf {\bibinfo {volume} {38}},\ \bibinfo {pages} {135004} (\bibinfo {year} {2021}{\natexlab{b}})},\ \Eprint {http://arxiv.org/abs/2101.07642} {arXiv:2101.07642 [gr-qc]} \BibitemShut {NoStop}%
\bibitem [{\citenamefont {van~der Marel}(1993)}]{1993ESOC...45...79V}%
  \BibitemOpen
  \bibfield  {author} {\bibinfo {author} {\bibfnamefont {R.~P.}\ \bibnamefont {van~der Marel}},\ }\href@noop {} {\ \textbf {\bibinfo {volume} {45}},\ \bibinfo {pages} {79} (\bibinfo {year} {1993})}\BibitemShut {NoStop}%
\bibitem [{\citenamefont {Fukushige}\ \emph {et~al.}(2004)\citenamefont {Fukushige}, \citenamefont {Kawai},\ and\ \citenamefont {Makino}}]{Fukushige:2003xc}%
  \BibitemOpen
  \bibfield  {author} {\bibinfo {author} {\bibfnamefont {T.}~\bibnamefont {Fukushige}}, \bibinfo {author} {\bibfnamefont {A.}~\bibnamefont {Kawai}}, \ and\ \bibinfo {author} {\bibfnamefont {J.}~\bibnamefont {Makino}},\ }\href {\doibase 10.1086/383192} {\bibfield  {journal} {\bibinfo  {journal} {Astrophys. J.}\ }\textbf {\bibinfo {volume} {606}},\ \bibinfo {pages} {625} (\bibinfo {year} {2004})},\ \Eprint {http://arxiv.org/abs/astro-ph/0306203} {arXiv:astro-ph/0306203} \BibitemShut {NoStop}%
\bibitem [{\citenamefont {Gebhardt}\ \emph {et~al.}(2000)\citenamefont {Gebhardt}, \citenamefont {Bender}, \citenamefont {Bower}, \citenamefont {Dressler}, \citenamefont {Faber}, \citenamefont {Filippenko}, \citenamefont {Green}, \citenamefont {Grillmair}, \citenamefont {Ho}, \citenamefont {Kormendy}, \citenamefont {Lauer}, \citenamefont {Magorrian}, \citenamefont {Pinkney}, \citenamefont {Richstone},\ and\ \citenamefont {Tremaine}}]{Gebhardt:2000fk}%
  \BibitemOpen
  \bibfield  {author} {\bibinfo {author} {\bibfnamefont {K.}~\bibnamefont {Gebhardt}}, \bibinfo {author} {\bibfnamefont {R.}~\bibnamefont {Bender}}, \bibinfo {author} {\bibfnamefont {G.}~\bibnamefont {Bower}}, \bibinfo {author} {\bibfnamefont {A.}~\bibnamefont {Dressler}}, \bibinfo {author} {\bibfnamefont {S.~M.}\ \bibnamefont {Faber}}, \bibinfo {author} {\bibfnamefont {A.~V.}\ \bibnamefont {Filippenko}}, \bibinfo {author} {\bibfnamefont {R.}~\bibnamefont {Green}}, \bibinfo {author} {\bibfnamefont {C.}~\bibnamefont {Grillmair}}, \bibinfo {author} {\bibfnamefont {L.~C.}\ \bibnamefont {Ho}}, \bibinfo {author} {\bibfnamefont {J.}~\bibnamefont {Kormendy}}, \bibinfo {author} {\bibfnamefont {T.~R.}\ \bibnamefont {Lauer}}, \bibinfo {author} {\bibfnamefont {J.}~\bibnamefont {Magorrian}}, \bibinfo {author} {\bibfnamefont {J.}~\bibnamefont {Pinkney}}, \bibinfo {author} {\bibfnamefont {D.}~\bibnamefont {Richstone}}, \ and\ \bibinfo {author} {\bibfnamefont {S.}~\bibnamefont {Tremaine}},\ }\href {\doibase 10.1086/312840}
  {\bibfield  {journal} {\bibinfo  {journal} {The Astrophysical Journal}\ }\textbf {\bibinfo {volume} {539}},\ \bibinfo {pages} {L13} (\bibinfo {year} {2000})}\BibitemShut {NoStop}%
\bibitem [{\citenamefont {Hernquist}(1990)}]{Hernquist:1990be}%
  \BibitemOpen
  \bibfield  {author} {\bibinfo {author} {\bibfnamefont {L.}~\bibnamefont {Hernquist}},\ }\href {\doibase 10.1086/168845} {\bibfield  {journal} {\bibinfo  {journal} {Astrophys. J.}\ }\textbf {\bibinfo {volume} {356}},\ \bibinfo {pages} {359} (\bibinfo {year} {1990})}\BibitemShut {NoStop}%
\bibitem [{\citenamefont {Tremaine}\ \emph {et~al.}(2002)\citenamefont {Tremaine}, \citenamefont {Gebhardt}, \citenamefont {Bender}, \citenamefont {Bower}, \citenamefont {Dressler}, \citenamefont {Faber}, \citenamefont {Filippenko}, \citenamefont {Green}, \citenamefont {Grillmair}, \citenamefont {Ho}, \citenamefont {Kormendy}, \citenamefont {Lauer}, \citenamefont {Magorrian}, \citenamefont {Pinkney},\ and\ \citenamefont {Richstone}}]{Tremaine:2002js}%
  \BibitemOpen
  \bibfield  {author} {\bibinfo {author} {\bibfnamefont {S.}~\bibnamefont {Tremaine}}, \bibinfo {author} {\bibfnamefont {K.}~\bibnamefont {Gebhardt}}, \bibinfo {author} {\bibfnamefont {R.}~\bibnamefont {Bender}}, \bibinfo {author} {\bibfnamefont {G.}~\bibnamefont {Bower}}, \bibinfo {author} {\bibfnamefont {A.}~\bibnamefont {Dressler}}, \bibinfo {author} {\bibfnamefont {S.~M.}\ \bibnamefont {Faber}}, \bibinfo {author} {\bibfnamefont {A.~V.}\ \bibnamefont {Filippenko}}, \bibinfo {author} {\bibfnamefont {R.}~\bibnamefont {Green}}, \bibinfo {author} {\bibfnamefont {C.}~\bibnamefont {Grillmair}}, \bibinfo {author} {\bibfnamefont {L.~C.}\ \bibnamefont {Ho}}, \bibinfo {author} {\bibfnamefont {J.}~\bibnamefont {Kormendy}}, \bibinfo {author} {\bibfnamefont {T.~R.}\ \bibnamefont {Lauer}}, \bibinfo {author} {\bibfnamefont {J.}~\bibnamefont {Magorrian}}, \bibinfo {author} {\bibfnamefont {J.}~\bibnamefont {Pinkney}}, \ and\ \bibinfo {author} {\bibfnamefont {D.}~\bibnamefont {Richstone}},\ }\href {\doibase 10.1086/341002}
  {\bibfield  {journal} {\bibinfo  {journal} {The Astrophysical Journal}\ }\textbf {\bibinfo {volume} {574}},\ \bibinfo {pages} {740} (\bibinfo {year} {2002})}\BibitemShut {NoStop}%
\bibitem [{\citenamefont {{Begeman}}(1989)}]{1989A&A...223...47B}%
  \BibitemOpen
  \bibfield  {author} {\bibinfo {author} {\bibfnamefont {K.~G.}\ \bibnamefont {{Begeman}}},\ }\href@noop {} {\bibfield  {journal} {\bibinfo  {journal} {\aap}\ }\textbf {\bibinfo {volume} {223}},\ \bibinfo {pages} {47} (\bibinfo {year} {1989})}\BibitemShut {NoStop}%
\bibitem [{\citenamefont {Milgrom}(1983{\natexlab{a}})}]{Milgrom:1983ca}%
  \BibitemOpen
  \bibfield  {author} {\bibinfo {author} {\bibfnamefont {M.}~\bibnamefont {Milgrom}},\ }\href {\doibase 10.1086/161130} {\bibfield  {journal} {\bibinfo  {journal} {Astrophys. J.}\ }\textbf {\bibinfo {volume} {270}},\ \bibinfo {pages} {365} (\bibinfo {year} {1983}{\natexlab{a}})}\BibitemShut {NoStop}%
\bibitem [{\citenamefont {Milgrom}(1983{\natexlab{b}})}]{Milgrom:1983pn}%
  \BibitemOpen
  \bibfield  {author} {\bibinfo {author} {\bibfnamefont {M.}~\bibnamefont {Milgrom}},\ }\href {\doibase 10.1086/161131} {\bibfield  {journal} {\bibinfo  {journal} {Astrophys. J.}\ }\textbf {\bibinfo {volume} {270}},\ \bibinfo {pages} {371} (\bibinfo {year} {1983}{\natexlab{b}})}\BibitemShut {NoStop}%
\bibitem [{\citenamefont {de~Martino}\ \emph {et~al.}(2020)\citenamefont {de~Martino}, \citenamefont {Chakrabarty}, \citenamefont {Cesare}, \citenamefont {Gallo}, \citenamefont {Ostorero},\ and\ \citenamefont {Diaferio}}]{deMartino:2020gfi}%
  \BibitemOpen
  \bibfield  {author} {\bibinfo {author} {\bibfnamefont {I.}~\bibnamefont {de~Martino}}, \bibinfo {author} {\bibfnamefont {S.~S.}\ \bibnamefont {Chakrabarty}}, \bibinfo {author} {\bibfnamefont {V.}~\bibnamefont {Cesare}}, \bibinfo {author} {\bibfnamefont {A.}~\bibnamefont {Gallo}}, \bibinfo {author} {\bibfnamefont {L.}~\bibnamefont {Ostorero}}, \ and\ \bibinfo {author} {\bibfnamefont {A.}~\bibnamefont {Diaferio}},\ }\href {\doibase 10.3390/universe6080107} {\bibfield  {journal} {\bibinfo  {journal} {Universe}\ }\textbf {\bibinfo {volume} {6}},\ \bibinfo {pages} {107} (\bibinfo {year} {2020})},\ \Eprint {http://arxiv.org/abs/2007.15539} {arXiv:2007.15539 [astro-ph.CO]} \BibitemShut {NoStop}%
\bibitem [{\citenamefont {McGaugh}(2020)}]{McGaugh:2020ppt}%
  \BibitemOpen
  \bibfield  {author} {\bibinfo {author} {\bibfnamefont {S.}~\bibnamefont {McGaugh}},\ }\href {\doibase 10.3390/galaxies8020035} {\bibfield  {journal} {\bibinfo  {journal} {Galaxies}\ }\textbf {\bibinfo {volume} {8}},\ \bibinfo {pages} {35} (\bibinfo {year} {2020})},\ \Eprint {http://arxiv.org/abs/2004.14402} {arXiv:2004.14402 [astro-ph.GA]} \BibitemShut {NoStop}%
\bibitem [{\citenamefont {Salucci}(2019)}]{Salucci:2018hqu}%
  \BibitemOpen
  \bibfield  {author} {\bibinfo {author} {\bibfnamefont {P.}~\bibnamefont {Salucci}},\ }\href {\doibase 10.1007/s00159-018-0113-1} {\bibfield  {journal} {\bibinfo  {journal} {Astron. Astrophys. Rev.}\ }\textbf {\bibinfo {volume} {27}},\ \bibinfo {pages} {2} (\bibinfo {year} {2019})},\ \Eprint {http://arxiv.org/abs/1811.08843} {arXiv:1811.08843 [astro-ph.GA]} \BibitemShut {NoStop}%
\bibitem [{\citenamefont {Navarro}\ \emph {et~al.}(1996)\citenamefont {Navarro}, \citenamefont {Frenk},\ and\ \citenamefont {White}}]{Navarro:1995iw}%
  \BibitemOpen
  \bibfield  {author} {\bibinfo {author} {\bibfnamefont {J.~F.}\ \bibnamefont {Navarro}}, \bibinfo {author} {\bibfnamefont {C.~S.}\ \bibnamefont {Frenk}}, \ and\ \bibinfo {author} {\bibfnamefont {S.~D.~M.}\ \bibnamefont {White}},\ }\href {\doibase 10.1086/177173} {\bibfield  {journal} {\bibinfo  {journal} {Astrophys. J.}\ }\textbf {\bibinfo {volume} {462}},\ \bibinfo {pages} {563} (\bibinfo {year} {1996})},\ \Eprint {http://arxiv.org/abs/astro-ph/9508025} {arXiv:astro-ph/9508025} \BibitemShut {NoStop}%
\bibitem [{\citenamefont {Herrera}\ and\ \citenamefont {Santos}(1997)}]{Herrera:1997plx}%
  \BibitemOpen
  \bibfield  {author} {\bibinfo {author} {\bibfnamefont {L.}~\bibnamefont {Herrera}}\ and\ \bibinfo {author} {\bibfnamefont {N.~O.}\ \bibnamefont {Santos}},\ }\href {\doibase 10.1016/S0370-1573(96)00042-7} {\bibfield  {journal} {\bibinfo  {journal} {Phys. Rept.}\ }\textbf {\bibinfo {volume} {286}},\ \bibinfo {pages} {53} (\bibinfo {year} {1997})}\BibitemShut {NoStop}%
\bibitem [{\citenamefont {Mak}\ and\ \citenamefont {Harko}(2003)}]{Mak:2001eb}%
  \BibitemOpen
  \bibfield  {author} {\bibinfo {author} {\bibfnamefont {M.~K.}\ \bibnamefont {Mak}}\ and\ \bibinfo {author} {\bibfnamefont {T.}~\bibnamefont {Harko}},\ }\href {\doibase 10.1098/rspa.2002.1014} {\bibfield  {journal} {\bibinfo  {journal} {Proc. Roy. Soc. Lond. A}\ }\textbf {\bibinfo {volume} {459}},\ \bibinfo {pages} {393} (\bibinfo {year} {2003})},\ \Eprint {http://arxiv.org/abs/gr-qc/0110103} {arXiv:gr-qc/0110103} \BibitemShut {NoStop}%
\bibitem [{\citenamefont {Cardoso}\ and\ \citenamefont {Pani}(2019)}]{Cardoso:2019rvt}%
  \BibitemOpen
  \bibfield  {author} {\bibinfo {author} {\bibfnamefont {V.}~\bibnamefont {Cardoso}}\ and\ \bibinfo {author} {\bibfnamefont {P.}~\bibnamefont {Pani}},\ }\href {\doibase 10.1007/s41114-019-0020-4} {\bibfield  {journal} {\bibinfo  {journal} {Living Rev. Rel.}\ }\textbf {\bibinfo {volume} {22}},\ \bibinfo {pages} {4} (\bibinfo {year} {2019})},\ \Eprint {http://arxiv.org/abs/1904.05363} {arXiv:1904.05363 [gr-qc]} \BibitemShut {NoStop}%
\bibitem [{\citenamefont {Raposo}\ \emph {et~al.}(2019)\citenamefont {Raposo}, \citenamefont {Pani}, \citenamefont {Bezares}, \citenamefont {Palenzuela},\ and\ \citenamefont {Cardoso}}]{Raposo:2018rjn}%
  \BibitemOpen
  \bibfield  {author} {\bibinfo {author} {\bibfnamefont {G.}~\bibnamefont {Raposo}}, \bibinfo {author} {\bibfnamefont {P.}~\bibnamefont {Pani}}, \bibinfo {author} {\bibfnamefont {M.}~\bibnamefont {Bezares}}, \bibinfo {author} {\bibfnamefont {C.}~\bibnamefont {Palenzuela}}, \ and\ \bibinfo {author} {\bibfnamefont {V.}~\bibnamefont {Cardoso}},\ }\href {\doibase 10.1103/PhysRevD.99.104072} {\bibfield  {journal} {\bibinfo  {journal} {Phys. Rev. D}\ }\textbf {\bibinfo {volume} {99}},\ \bibinfo {pages} {104072} (\bibinfo {year} {2019})},\ \Eprint {http://arxiv.org/abs/1811.07917} {arXiv:1811.07917 [gr-qc]} \BibitemShut {NoStop}%
\bibitem [{\citenamefont {Becerra}\ \emph {et~al.}(2024)\citenamefont {Becerra}, \citenamefont {Becerra-Vergara},\ and\ \citenamefont {Lora-Clavijo}}]{Becerra:2024wku}%
  \BibitemOpen
  \bibfield  {author} {\bibinfo {author} {\bibfnamefont {L.~M.}\ \bibnamefont {Becerra}}, \bibinfo {author} {\bibfnamefont {E.~A.}\ \bibnamefont {Becerra-Vergara}}, \ and\ \bibinfo {author} {\bibfnamefont {F.~D.}\ \bibnamefont {Lora-Clavijo}},\ }\href {\doibase 10.1103/PhysRevD.109.043025} {\bibfield  {journal} {\bibinfo  {journal} {Phys. Rev. D}\ }\textbf {\bibinfo {volume} {109}},\ \bibinfo {pages} {043025} (\bibinfo {year} {2024})},\ \Eprint {http://arxiv.org/abs/2401.10311} {arXiv:2401.10311 [astro-ph.HE]} \BibitemShut {NoStop}%
\end{thebibliography}%

\end{document}